\documentclass[a4paper,fleqn]{cas-sc}

\usepackage[authoryear,longnamesfirst]{natbib}
\usepackage{comment}
\def\tsc#1{\csdef{#1}{\textsc{\lowercase{#1}}\xspace}}
\tsc{WGM}
\tsc{QE}
\usepackage{subcaption}
\usepackage{siunitx}
\begin{document}
\let\WriteBookmarks\relax
\def\floatpagepagefraction{1}
\def\textpagefraction{.001}

\shorttitle{2pcf from DESI’s BGS}    

\shortauthors{}  

\title [mode = title]{Angular BAO Measurements with the DESI DR1 BGS Sample}  

\tnotemark[1] 

\tnotetext[1]{} 


\author[1,2,3]{Paula S. Ferreira}[orcid=0000-0002-7540-040X]

\cormark[1]

\fnmark[1]

\ead{ferreira_paula@zju.edu.cn}

\credit{Writing – review \& editing, Visualization, Validation, Supervision, Methodology, Interpretation, Formal analysis, Conceptualization}

\affiliation[1]{organization={Center for Cosmology and Computational Astrophysics, Institute for Advanced Study in Physics},
            addressline={Zhejiang University}, 
            city={Hangzhou},
%          citysep={}, % Uncomment if no comma needed between city and postcode
            postcode={310058}, 
            state={Zhejiang},
            country={China}}

\affiliation[2]{organization={Institute of Astronomy, School of Physics, Zhejiang University},
            addressline={Zhejiang University}, 
            city={Hangzhou},
%          citysep={}, % Uncomment if no comma needed between city and postcode
            postcode={310058}, 
            state={Zhejiang},
            country={China}}

\affiliation[3]{organization={Centro Brasileiro de Pesquisas Físicas},
            addressline={Rua Dr. Xavier Sigaud, 150}, 
            city={Rio de Janeiro},
%          citysep={}, % Uncomment if no comma needed between city and postcode
            postcode={22290-180}, 
            state={RJ},
            country={Brazil}}
\affiliation[4]{organization={Observatório Nacional},
            addressline={ Rua General José Cristino,77, São Cristóvão}, 
            city={Rio de Janeiro},
%          citysep={}, % Uncomment if no comma needed between city and postcode
            postcode={20921-400}, 
            state={RJ},
            country={Brazil}}
%Observat\'orio do Valongo, Universidade Federal do Rio de Janeiro, 20080-090, Rio de Janeiro, RJ, Brazil
\affiliation[5]{organization={Observat\'orio do Valongo, Universidade Federal do Rio de Janeiro},
            addressline={Ladeira Pedro Antônio, 43}, 
            city={Rio de Janeiro},
%          citysep={}, % Uncomment if no comma needed between city and postcode
            postcode={20080-090}, 
            state={RJ},
            country={Brazil}}

\author[3]{Ulisses Ribeiro}%

% Footnote of the second author
\fnmark[2]

% Email id of the second author

\credit{Validation and cosmological inference}

\author[1,2,3,5]{Pedro da Silveira Ferreira}[orcid=0000-0003-4936-0069]

% Footnote of the second author
\fnmark[3]

% Email id of the second author

\credit{Writing – review \& editing, interpretation of the results and random catalog generation}

\author[3]{Clécio R. Bom}[orcid=0000-0003-4383-2969]

% Footnote of the second author
\fnmark[4]
\credit{Review \& editing}
% Email id of the second author

\author[4]{Armando Bernui}[orcid=0000-0003-3034-0762]

% Footnote of the second author
\fnmark[5]
\credit{Writing – Review \& editing}

\cortext[1]{Corresponding author}



\begin{abstract}
We employ a model-independent approach in both the correlation function estimation and the angular BAO feature estimation by computing the angular two-point correlation function. First, we conducted a series of tests to the available DESI tracers to check their representativeness to angular clustering; the result was that, considering the completeness of the first data release across the footprint, we could only make use of the BGS sample for the effective redshifts 0.21 (BGS1) and 0.25 (BGS2). For a reliable analysis in such low redshift, we use the bootstrap of the data itself to construct a covariance matrix that accounts for systematics. We use a purely statistical method to correct the projection effects and find that our results show reasonable agreement with the $\theta_{\rm BAO}$ expected by the CPL parameters obtained by DESI DR1, being BGS1 $12.77 \pm 1.15$ degrees and BGS2 $11.70 \pm 1.21$ degrees. This means a tension at the $2.4\sigma$ ($2.7\sigma$) level for BGS1 (BGS2) CPL parametrization, while a $2.84\sigma$ ($3.02\sigma$) discrepancy within the predicted by $\Lambda$CDM. We conclude that, with the current sample available, the use of an angular correlation function serving as the BAO probe, although prefers the CPL parametrization, does not provide conclusive results regarding the best cosmological model.
\end{abstract}


\begin{comment}
 \begin{highlights}
\end{highlights}   
\end{comment}

% Keywords

\maketitle

\section{Introduction}

%The precision cosmology era is undergoing a revolution, where the search for the universe dynamics, including the time evolution of the elements that compose it, are being investigated with improved automatized instrumentation in the diverse --ongoing and in commissioning-- astronomical survey projects \cite{adame2024desi,splus}. 

The Baryon Acoustic Oscillations (BAO) scale is imprinted on the large-scale structure of the universe from sound waves propagating in the primordial plasma. 
After recombination, these waves froze, leaving a characteristic scale, a standard ruler, of approximately 150 Mpc in the comoving frame. 
This physical scale provides a powerful geometric probe for measuring the expansion history of the universe and constraining the properties of dark energy.

Computing the two-point correlation function (2PCF) of the distribution of cosmic objects, the BAO feature appears as a peak~\citep{Eisenstein05}. 
%Therefore, the distribution of cosmic objects in the observed universe has an excess of probability to grow at the location of the spherical shell. 
Alternatively, one can measure the Fourier-space equivalent of the 2PCF, that is, the power spectrum, where the BAO feature appears as a series of oscillations~\citep{Cole05}. 

While the 3-dimensional (3D) analysis combines radial and transverse information to measure the Hubble parameter $H(z)$ and the angular diameter distance $D_A(z)$ simultaneously, the transverse BAO signal offers a complementary and robust approach, where the BAO signature is expected as circular rings on the celestial sphere, with angular radius $\theta_{\rm BAO}(z)$. 

As in the BAO-3D studies, the transverse BAO feature was also detected in the analysis of diverse cosmic tracers and at different epochs of universe evolution \citep{Cole_2005,Percival_2010,Alam_2017,Gil_Mar_n_2020,Chan_2022,Adame_2025}.
These studies not only confirm the BAO as a universal phenomenon, but, more interestingly, provide measurements of this standard ruler at several redshifts. 

The angular BAO measurement offers several distinct advantages that make it a powerful complement to its 3D counterpart. First, the transverse BAO signal is independent of redshift-space distortions, which arise from peculiar velocities along the line of sight and can significantly complicate the interpretation of 3D clustering measurements. By working in angular projection, the transverse BAO provides a cleaner measurement of the angular diameter distance \(D_A(z)\), as the clustering signal is integrated over the line of sight, effectively averaging out radial peculiar velocities. The angular BAO scale is given by \(\theta_{\rm BAO}(z) = r_s(z_d) / [(1+z)D_A(z)]\), where \(r_s(z_d)\) is the comoving sound horizon at the drag epoch and \(D_A(z)\) is the angular diameter distance. Second, the angular approach is particularly well suited for photometric surveys, where accurate spectroscopic redshifts are unavailable. In such surveys, the 2D transverse BAO can be extracted with a statistical precision comparable to that of 3D analyses. The angular correlation function is obtained by projecting the 3D correlation function \(\xi(r)\) along the line of sight: \(w(\theta) = \int dz \, \phi(z)^2 \int d\pi \, \xi(\sqrt{r^2 + \pi^2})\), where \(\phi(z)\) is the normalised redshift distribution of the sample and \(\pi\) is the line-of-sight separation. Third, the transverse BAO scale is directly proportional to the sound horizon at the drag epoch, offering a model-independent probe of cosmic expansion when combined with measurements of the Hubble parameter \(H(z)\) from other probes, since \(D_A(z) = (1+z)^{-1} \int_0^z dz' / H(z')\). Fourth, because the angular correlation function does not require knowledge of the radial distance, it is less sensitive to systematic errors in redshift estimation and can serve as a cross-check for systematic biases that might affect 3D BAO measurements. Finally, stacking angular BAO measurements across multiple redshift bins can provide a robust constraint on the cosmological distance-redshift relation, independent of the specific BAO model.

It is worth noting, however, that the reliability of such angular measurements relies critically on survey volumes that are sufficiently large to suppress cosmic variance and window-function effects. This condition becomes particularly important at low redshifts, where the BAO angular scale can be comparable to the transverse size of the survey footprint itself, potentially compromising the interpretation of the detected signal.

The concordance model of cosmology, 
$\Lambda$CDM, is being exhaustively tested with precise data recently released by the Dark Energy Spectroscopic Instrument (DESI) Collaboration \citep{desicollaboration2025datarelease1dark,Anonymous_2025,desicollaboration2025desidr2resultsii}. 
In fact, the focus of this large astronomical survey is to obtain precise measurements of the BAO scale, by analyzing a series of cosmic tracers at different redshifts. 
%Such measurements were recently analysed by the DESI collaboration, pointing out 
Recently, statistical analyses of the DESI collaboration combined these BAO measurements with Planck CMB and supernovae type Ia data, concluding that the flat-$\Lambda$CDM model is not the preferred model \citep{desicollaboration2025desidr2resultsii} with $w_0<-1$ and $w_a\neq 0$. Reasons and possible explanations for these tensions are thoroughly discussed in \cite{Di_Valentino_2025}. 
Instead, the data favoured the $\omega_0 \omega_a$CDM model, with evolving dark energy, considering the CPL parametrisation \citep{CHEVALLIER_2001, Linder_2003}. 

\begin{table*}[t]
\centering
\footnotesize
\setlength{\tabcolsep}{4pt} % Reduce column padding
\begin{tabular}{@{}lrrrrrrrr@{}}
\toprule
 & \multicolumn{2}{c}{BGS} & \multicolumn{2}{c}{LRG} & \multicolumn{2}{c}{ELG} & \multicolumn{2}{c}{QSO} \\
\cmidrule(lr){2-3}\cmidrule(lr){4-5}\cmidrule(lr){6-7}\cmidrule(lr){8-9}
 & SGC & NGC & SGC & NGC & SGC & NGC & SGC & NGC \\
\midrule
Area (deg$^2$) & 3,035 & 6,829 & 3,497 & 5,934 & 3,436 & 5,845 & 3,473 & 5,875 \\
Galaxies ($\times10^3$) & 2,910 & 1,048 & 662 & 1,476 & 611 & 1,821 & 430 & 793 \\
Density (deg$^{-2}$) & 345 & 426 & 189 & 249 & 178 & 312 & 124 & 135 \\
\bottomrule
\end{tabular}
\caption{Sky coverage by tracer type, showing area, galaxy counts, and number density. SGC and NGC refer to the Southern and Northern Galactic Caps respectively.}
\label{tab:sky}
\end{table*}

The goal of this work is to analyse the transverse BAO signal in low-redshift bins (low-z), using \cite{sanchez2011tracing} polynomial fit to obtain cosmological constraints and check whether DESI DR1 shows model-free BAO results for different fiducial cosmologies analysed by \cite{desicollaboration2025desidr2resultsii} projection effects.

%%---------------------------------------------------------
\section{Data set}

In this study, we use data from the first release (DR1)\footnote{\url{https://data.desi.lbl.gov/public/dr1/survey/catalogs/dr1/LSS/iron/LSScats/v1.5/}} by the DESI, a state-of-the-art spectroscopic instrument located at the 4-meter Mayall Telescope at Kitt Peak National Observatory in Arizona, USA \citep{levi2019darkenergyspectroscopicinstrument}. DESI is now conducting a massive spectroscopic survey, mapping over 14,000 square degrees of the sky to measure the redshifts of tens of millions of galaxies and quasars. We assessed suitable tracers for angular clustering and determined, after several tests, that only the low-redshift galaxy samples are adequate. One of the biggest problems using DR1 is the inhomogeneity of completeness across the footprint, which is a major drawback considering an angular-based measurement.

\subsection{The DESI Tracers}

The Dark Energy Spectroscopic Instrument (DESI) is a five-year survey designed to map the large-scale structure of the Universe by collecting spectra of approximately 40 million extragalactic objects across more than 14,000 square degrees of the sky. To probe the cosmic expansion history across a broad redshift range, the DESI survey uses four complementary extragalactic tracers: the Bright Galaxy Survey (BGS), Luminous Red Galaxies (LRGs), Emission Line Galaxies (ELGs), and Quasars (QSOs). These tracers are photometrically selected from the DESI Legacy Imaging Surveys and cover different redshift intervals, allowing DESI to span approximately 80\% of cosmic history. 

The Bright Galaxy Survey (BGS) targets low-redshift galaxies (\(0.05<z<0.4\)), with a median redshift of approximately \(z\approx 0.2\). It is selected to be a nearly magnitude-limited sample and will provide the most precise BAO and redshift-space distortion (RSD) measurements at low redshift. The BGS is divided into Bright and Faint components, with target densities of about 800 and 600 objects per square degree, respectively. Its target selection and validation are described in detail by \citet{Hahn2023}.

The Luminous Red Galaxy (LRG) sample covers the intermediate redshift range \(0.4<z\lesssim 1.0\). With a target density of 605 deg\(^{-2}\) and a comoving number density of \(5\times 10^{-4}\,h^{3}\,\text{Mpc}^{-3}\) in \(0.4<z<0.8\), it represents a significant improvement in density over previous LRG surveys such as SDSS, BOSS, and eBOSS. The sample is selected using \(g\), \(r\), \(z\), and W1 photometry and exhibits a high spectroscopic success rate, with 98.9\% of observed targets yielding confident redshifts. The selection and performance of the DESI LRG sample are presented by \citet{Zhou2023}.

Emission Line Galaxies (ELGs) are the most numerous DESI tracer, comprising about one-third of all targets. They probe the redshift range \(0.6<z<1.6\), with the highest-redshift portion (\(1.1<z<1.6\)) expected to provide the tightest cosmological constraints. The ELG target selection relies on a \(g\)-band magnitude cut and a \((g-r)\) versus \((r-z)\) colour box. The main sample is composed of two disjoint subsamples with target densities of approximately 1940 and 460 deg\(^{-2}\), respectively. The target selection and validation of DESI ELGs are detailed by \citet{Raichoor2023}.

Finally, the Quasar (QSO) sample extends the redshift coverage up to \(z=2.1\). Quasars are selected using a machine-learning approach, as their colours are similar to those of stars. The QSO sample serves a dual role: at \(0.9<z<2.1\), quasars are used as direct tracers of large-scale structure. The target selection for the QSO sample is presented by \citet{Chaussidon2023}.

\subsection{On the selection of the best samples}
%We had to test the number density for all available tracers from DESI. 
The inspection conducted for each tracer involved the assessment of three key parameters: 
first, the total count of objects; 
second, the fraction of sky covered;  
third, the density of galaxies per square degree per redshift bin for the Southern and Northern Galactic Caps. We applied a redshift bin width of $\Delta z = 0.1$ specifically for LRGs, ELGs, and QSOs, see Table~\ref{tab:sky}. However, this did not reveal any notable BAO features. Upon exploring more refined binning strategies, we discovered that the implementation of highly granular bin sizes proved impractical for any of the tracers, particularly within the context of a conventional analysis framework like the one introduced in this study. This observation, quite naturally, prompted us to address the challenge of identifying the most effective bin width. See the discarded bins in Appendix~\ref{ap:discard}.

Using the cosmological model $\Lambda$CDM, which is a good enough approximation for the selection tests, we conducted an estimation of the angular scale associated with the BAO signal within a circular region, examining different bin widths to assess its impact. Subsequently, we identified the specific data bins that would enclose such regions, ensuring that each contained a minimum of 500 galaxies. With these bins selected, the subsequent phase involves applying tests on the angular two-point correlation function (2PCF) to investigate the structure within the data. 
From this analysis, the remaining candidate bins are written in the Table \ref{tab:data_bins}, where 
\begin{equation}
z_{\rm eff} = \frac{\sum_{i<j} \rho_i \rho_j(z_i+z_j)}{2\sum_{i<j}\rho_i \rho_j}\,,  
\end{equation}with $i$ and $j$ being the indices of each galaxy and $\rho$ the \texttt{WEIGHT\_SYS} \citep{Ross_2025}.

\subsection{Bins choice}
\begin{table}
    \centering
    \begin{tabular}{lccc}
        \toprule

        Tracer & $z_{\rm eff}$ & bin &  number of objects \\
        \hline
        &&&\vspace{-3mm}\\
        BGS NGC3 & 0.27 &  $0.26\leq z<0.28$ & 168,440\\
         BGS NGC4 & 0.31 &  $0.30\leq z<0.32$& 120,925\\
         BGS NGC5 & 0.33 & $0.32\leq z<0.34$& 104,837\\
         BGS SGC1 & 0.25 & $0.24\leq z<0.26$ & 65,461\\
         BGS SGC2 & 0.27 & $0.26\leq z<0.28$ & 60,372\\
         BGS SGC3 & 0.29 & $0.28\leq z<0.30$& 53,020\\
         QSO NGC & 1.52 & $1.00\leq z<2.00$ & 458,790\\
         LRG NGC & 0.82 & $0.70\leq z<0.95$ & 714,089 \\
         \hline
         BGS NGC1 & 0.21 & $0.20\leq z<0.22$ & 212,411\\
         BGS NGC2 & 0.25 & $0.24\leq z<0.26$ & 179,202\\
        \bottomrule
    \end{tabular}
    \caption{Bins with enough clustered galaxies that could contain the BAO feature. The samples we suited to this analysis are in the bottom of the table.}
    \label{tab:data_bins}
\end{table}

Regarding our sample selection, several methodological decisions warrant clarification. First, due to the DESI fibre-collision systematics, any angular clustering measurement must consider scales larger than approximately 0.05 degrees \citep{pinon2025mitigation}. This angular limit guided our choice of binning strategy.

From the selection criteria described above, the BGS NGC1 and NGC2 samples yielded the most reliable angular correlation functions. Our binning choices are broadly consistent with the analysis by \cite{novell2025full}, who opted to drop the ELG sample from their 3D clustering analysis owing to significant systematics that are better handled by the DESI Collaboration's official pipeline. The only distinction between our approach and theirs is that we additionally discarded the QSO sample, as the resulting angular bins would suffer considerably from projection effects, which are particularly severe for sparse, high-redshift tracers.

Two further points are important regarding the BGS sample. We deliberately chose not to apply additional cuts to the survey mask to make it more uniform, for two reasons. First, such a procedure would render our analysis less directly comparable with the official DESI Collaboration results, which we consider an important benchmark. Second, and more critically, cutting a rectangular region from the mask could introduce a biased analysis, as any remaining structures within that particular region would be artificially enhanced, especially given the high density and low redshift of the BGS sample, where local cosmic variance is already substantial.

Finally, we do not include the lowest redshift bin from the BGS sample in our analysis. The reason is that, at such shallow redshifts, non-linear structure formation becomes significant, and accurately modelling the angular clustering requires realistic mock catalogues that can reproduce these local non-linearities. As emphasised by \cite{Novell-Masot_2025}, a reliable fit to the angular correlation function at low redshifts demands a representative model that accounts for these effects. Since our current mock set does not adequately capture this regime, we conservatively exclude that redshift range to avoid introducing uncontrolled systematics into our measurements.

\begin{table}[t]
    \centering
    \begin{tabular}{c}
    \toprule
        Mock realization parameters\\
        \hline
         $N_{grid}=1024$\\
         $N_{side} = 1024$\\
         Tracer Kernel: $\Theta(z<0.1)$\\
         $\Omega_m$ :$ 0.3153$\\
         $\Omega_\Lambda$: $0.6847$\\
         $w$: $-1.0$\\
         $n_s$: $0.9649$\\
         $\sigma_8$: $0.8111$\\
         Transfer funtion: \texttt{eiseinteins\_hu}\\
         Galaxy bias: $b(z)=\sqrt{1+z}$\\
    \bottomrule
    \end{tabular}
    \caption{\texttt{CoLoRe} parameters used for mock realizations.}
    \label{tab:mocks_params}
\end{table}

\subsection{Mocks}
Because the BGS sample is susceptible to local inhomogeneities, we must make an effort to construct mock realisations that take these inhomogeneities into account. DESI Collaboration has carefully studied the ideal mock realisations to their samples and, as previously done by SDSS, has chosen to use 100 realisations from \texttt{EZmocks} \citep{chuang2015ezmocks}  based on improvements to the Z'eldovich approximation (ZA). The ZA defines a Lagrange-space regime that leads to quasi-nonlinear solution, which breaks beyond quasi-nonlinear due to caustics occurrence.

Here, we will use \texttt{CoLoRe} \citep{ramirez2022colore} to produce mocks using Lagrangian Perturbation Theory (LPT) whose description has intrinsically non-linear information \citep{bernardeau2002large}. We chose to use first-order LPT (LPT1). In \cite{ramirez2022colore}, \texttt{CoLoRe} has tested its potential to generate stage-IV survey simulations that include the BGS sample. 

\begin{figure}[H]
    \centering

    \includegraphics[width=.8\linewidth]{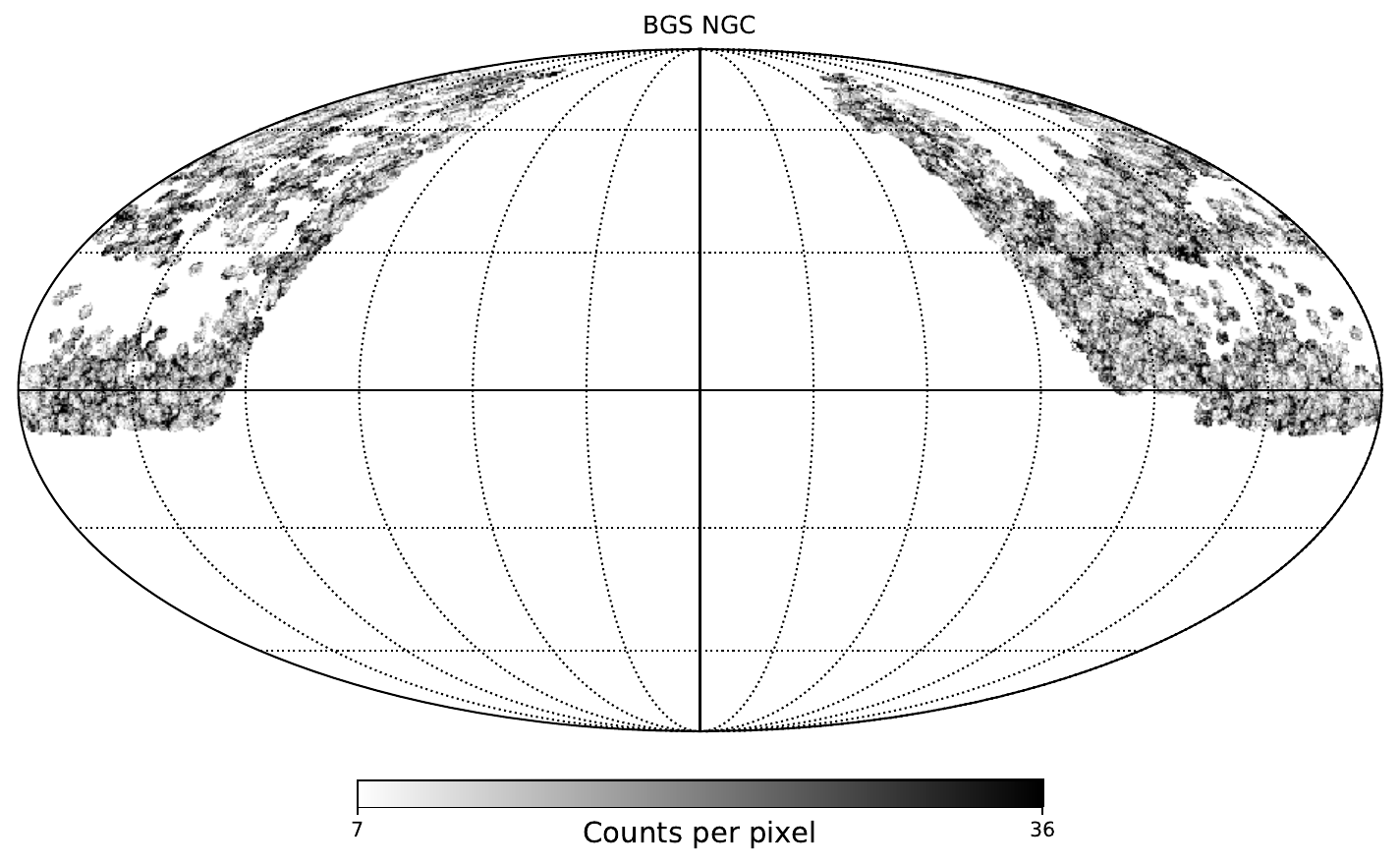}
    \caption{BGS example mock. The darker regions indicate the most populated pixels.}
    \label{fig:bgsmock}

\end{figure}

The fiducial cosmology chosen to construct the mocks is based on Planck 18 results \cite{planck18}; see Table \ref{tab:mocks_params} for $1,000$ realisations for each tracer. We use a single random catalogue for the realisations, like the DESI Collaboration, these randoms have $\sim 82$ times the number of the mock galaxies. Fig.~\ref{fig:bgsmock} shows an example of a mock realisation for the BGS NGC samples. They have exactly the sky footprint of the real data. \texttt{CoLoRe} operates at the field level by first generating a density field on a three-dimensional grid using first-order Lagrangian perturbation theory (LPT1). The code begins by computing the linear matter power spectrum for the input cosmology using the Eisenstein \& Hu transfer function \cite{eisenstein_hu} and then generates a Gaussian random field with the corresponding power spectrum. From this Gaussian field, CoLoRe computes the displacement field using the Z'eldovich approximation, which provides a more accurate description of the large-scale structure than a simple lognormal model while remaining computationally efficient. The displacement field is used to move particles from their initial Lagrangian positions to their final Eulerian positions, producing a realistic matter density field that captures the BAO feature and the correct phase information.

The galaxy bias is incorporated in a redshift-dependent manner following \(b(z) = \sqrt{1+z}\), consistent with the expected scale-independent bias for the BGS sample. The mock catalogues are generated on a grid with \(N_{\rm grid}=1024\) and \(N_{\rm side}=1024\) for the \texttt{healpix} projection, ensuring sufficient spatial resolution to reproduce the angular clustering signal down to the scales of interest. Importantly, the mocks are constructed to have exactly the same sky footprint as the real DESI data, including all geometric and observational selection effects. This is achieved by applying the same angular mask and completeness weights to the mock catalogues as those used for the real data.

The choice for $\Lambda$CDM fiducial cosmology is not in violation of the results obtained by the DESI Collaboration that favour the CPL model. The conditions for what is supposed to be a fair sample for an ensemble of universes were first postulated by \cite{peebles1973statistical}. Our objective is to determine the size of the BAO peak. Therefore, a precise match between mock and observed data is unnecessary; we only need to capture the same statistical distribution. It is essential to note that we are analysing the 2D correlation function, excluding the 3D component, and without relying on a fiducial cosmology for distance relation. Our primary concern is ensuring that the mocks fall within the correct mask and have sufficient number density, a robustness test is described in Appendix~\ref{ap:robust_mock}.

The number of mocks followed the suggested rule of thumb by \cite{Hartlap_2006}, where the ratio 
\begin{equation*}
    \frac{N_{mocks}-N_{data \, points}-2}{N_{mocks}-1},
\end{equation*}
should not be small ($\ll1$) in order to find good constraints.

\begin{figure*}[ht]
\centering
\begin{subfigure}[t]{0.48\textwidth}
  \centering
  \includegraphics[width=\linewidth]{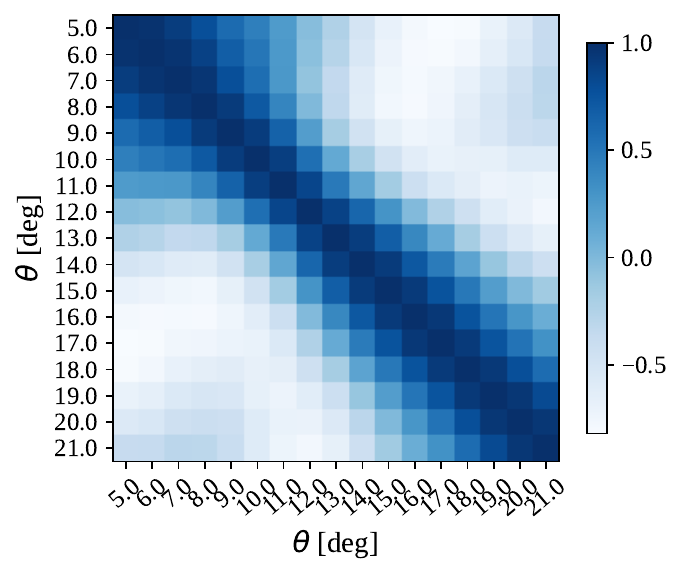}
  \caption{BGS1}\label{fig:bgs_cov1}
\end{subfigure}\hfill
\begin{subfigure}[t]{0.48\textwidth}
  \centering
  \includegraphics[width=\linewidth]{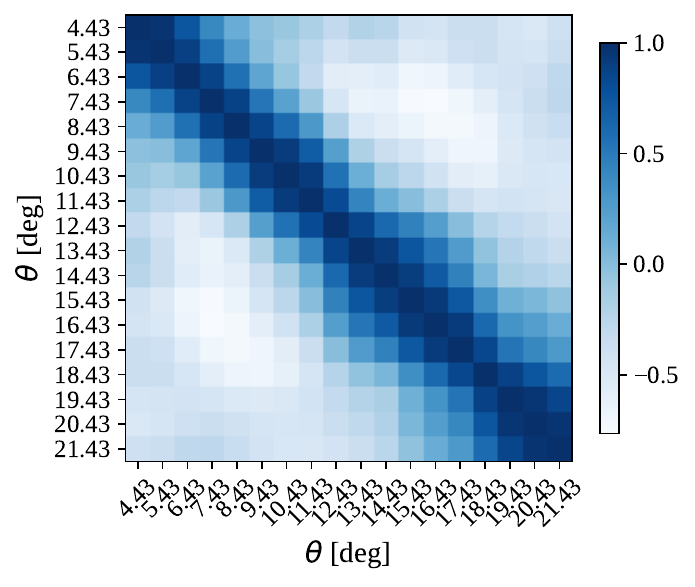}
  \caption{BGS2}\label{fig:bgs_cov2}
\end{subfigure}
\caption{Correlation matrix from the covariance of the 1,000 mock 2PCFs for BGS1 and BGS2.}
\label{fig:combined_cov1}
\end{figure*}

\begin{figure*}[ht]
\centering
\begin{subfigure}[t]{0.48\textwidth}
  \centering
  \includegraphics[width=\linewidth]{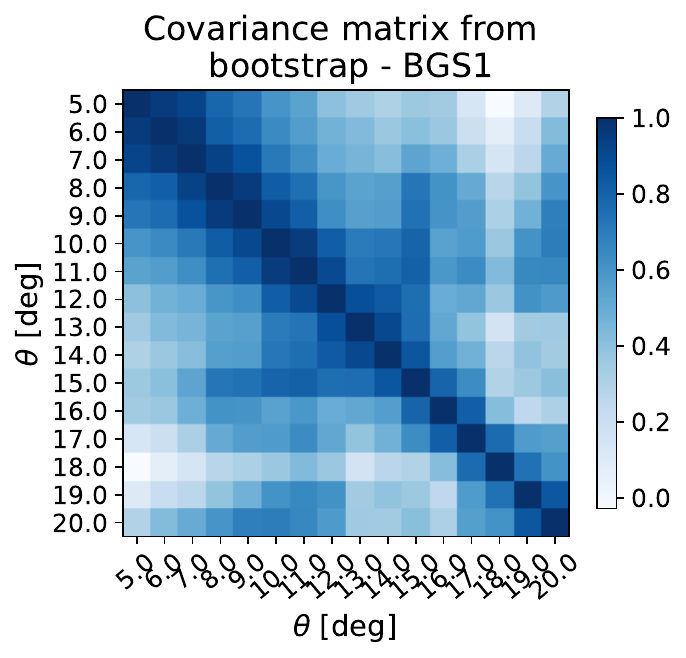}
  \caption{BGS1}\label{fig:bgs_cov_b1}
\end{subfigure}\hfill
\begin{subfigure}[t]{0.48\textwidth}
  \centering
  \includegraphics[width=\linewidth]{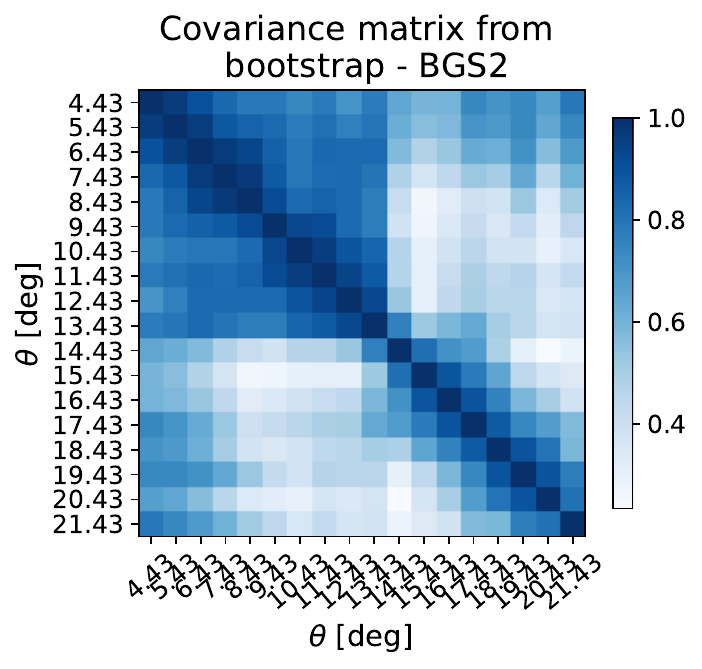}
  \caption{BGS2}\label{fig:bgs_cov_b2}
\end{subfigure}
\caption{Correlation matrix from the covariance of the 100 bootstrap realisations 2PCFs for BGS1 and BGS2.}
\label{fig:combined_cov2}
\end{figure*}

For a realistic pair counting, we estimated the mocks' weights using a Nearest-Neighbour interpolation with the observed data's right ascension, declination and redshift as the input variables to find the \texttt{WEIGHT\_SYS}.

\subsection{Random catalogs}

We construct random catalogs with a density 100 times larger than that of the data for the BGS (NGC) sample. The survey footprint is represented as the set of \texttt{HEALPix} pixel IDs at \({\rm NSIDE}=2^{29}\). At this resolution, the full sky is a discrete grid of \(N_{\rm pix}\simeq 3.46\times10^{18}\) pixels, hence the BGS \((f_{\rm sky}\simeq0.16)\) footprint corresponds to \(\simeq5.6\times10^{17}\) admissible pixel IDs. For each random object, we sample uniformly from this grid of possible positions and assign the direction of the corresponding pixel center; this ID-based sampling avoids enumerating the full grid and is computationally very efficient. We then draw a redshift from the empirical \(N(z)\) of the matching sample (or tomographic bin), so that the randoms reproduce the observed radial selection.

By construction, these randoms are isotropic within the angular mask, as expected for a \texttt{HEALPix} grid map, and match the data’s redshift distribution, while the \(100\times\) larger size suppresses shot noise in pair counts. The extremely fine NSIDE makes any residual pixelization negligible. Nevertheless, our implementation can randomly perturb positions to remove any center-of-pixel imprint. At the grid resolution used, each pixel has a characteristic angular radius of \(\sim 2\times10^{-4}\) arcsec, and we verified that such a displacement has no measurable impact. The same procedure is in all correlation-function measurements.

\section{Methodology}
\subsection{Estimator}\label{sec:methods}

Utilizing the counts of pairs, the angular two-point correlation function for each redshift bin was calculated through the application of the Landy-Szalay estimator \citep{landy1993bias}. For each angular function, normalization was carried out corresponding to the respective i-th tracer. The expression for the angular correlation function, denoted as $\mathrm{w}_i$, pertaining to the i-th bin, is represented by the equation provided below 
\begin{equation}
\mathrm{w}_i(\theta)= \left( \frac{N_{i,rand}}{N_{i,data}}\right)^2 \frac{DD_i(\theta)}{RR_i(\theta)} - 2 \frac{N_{i,rand}}{N_{i,data}}\frac{DR_i(\theta)}{RR_i(\theta)}+1 \text{ ,}
\label{eq:tpcf}
\end{equation}
where $N_{i,rand}$ is the number of galaxies in the random catalog, and $N_{i,data}$ in a tracer. The counting is based on \texttt{WEIGHT\_SYS} \citep{Ross_2025}, we are not using Feldmann-Kaiser-Peacock's \citep{fkp94} weight becase it is model-dependent and only affects the 3D clustering, our analysis is purely angular.

To substantiate the reliability of our methodology, we require the acquisition of survey catalogue realisations derived from simulated catalogues. These mocks serve as an approximate representation of a survey, paying particular attention to its spatial coverage and redshift distribution. It is crucial to acknowledge that during the analysis of clustering statistics, one major challenge posed by these mocks is the accurate representation of nonlinear effects, which become unavoidable at lower redshift values as highlighted by previous studies. 

In our study, we employ mocks to calculate $\mathrm{w}_k(\theta)$ for the $k$-th mock, enabling us to rigorously evaluate the proposed method. At this juncture, the covariance matrix $C_{ml}$ becomes an integral component:
\begin{align}
    C_{ml} = \frac{n_{\text{eff}}^{-1}}{N-1} \displaystyle \sum\limits_{k=1}^{N=1000} &[ \mathrm{w}_k(\theta_m) - \mathrm{\bar{w}}(\theta_m)] \nonumber \times \\
    &\,\![\mathrm{w}_{k}(\theta_l)-\mathrm{\bar{w}}(\theta_l)],\label{eq:covmocks}
\end{align}
where $\mathrm{\bar{w}}$ is the average over the correlation function ${\rm w}_k(\theta)$ of each mock. The covariance matrices can be found in Fig.\ref{fig:combined_cov1} and \ref{fig:combined_cov2}, all three sets are showing a strong diagonal structure with minimal off-diagonal signals.

As our principal strategy for quantifying the final $\theta_{BAO}$, we employ a bootstrap resampling scheme. Specifically, we generate 100 bootstrap realisations of the data set. In each realisation, we draw, with replacement, a subset corresponding to 80\% of the original survey area (or, equivalently, 80\% of the galaxy catalogue), in such a way that both the surveyed sky fraction and the observed redshift distribution are preserved. An identical resampling procedure is applied to the random catalogues that enter the estimation of the angular correlation function, ensuring that the data–random pairing remains self‑consistent in every bootstrap sample. The covariance matrix from the bootstrap realisations is then computed as

\begin{align}
C_{ml}^{\text{BS}} = \frac{1}{N_B - 1} \sum_{k=1}^{N_B = 100} & \big[ w_k^{\text{BS}}(\theta_m) - \bar{w}^{\text{BS}}(\theta_m) \big] \times \big[ w_k^{\text{BS}}(\theta_l) - \bar{w}^{\text{BS}}(\theta_l) \big],
\label{eq:covbootstrap}
\end{align}
where $w_k^{BS}(\theta)$ is the correlation function measured from the k-th bootstrap realisation and $\bar{w}^{BS}(\theta)$ is the average over all bootstrap realisations

This bootstrap framework offers two key benefits. First, it is fully empirical and model‑agnostic: the resulting uncertainty estimates depend exclusively on the measured data and do not require any assumptions about the form of the underlying cosmological model or the power spectrum. Second, it automatically incorporates the exact survey footprint, fibre‑collision effects, non‑uniform completeness, and any remaining observational systematics that may not be perfectly represented in idealised mock catalogues or analytic calculations. Consequently, the covariance matrices inferred from the bootstrap procedure are intrinsically tailored to the real survey, including its geometry and observational complexities.

\subsection{Model}\label{sec:model}
We used \cite{sanchez2011tracing} expression below for the angular two-point correlation function to fit our estimated function 
%${\rm w}(\theta)$ 
\begin{equation}\label{eq:sanchez_model}
{\rm w}(\theta) = A+B\theta^\gamma+C e^{-\frac{(\theta-\theta_{fit})^2}{2\sigma^2}}\,.
\end{equation}

\begin{figure*}[ht]
\centering
\begin{subfigure}[t]{0.48\textwidth}
  \centering
  \includegraphics[width=\linewidth]{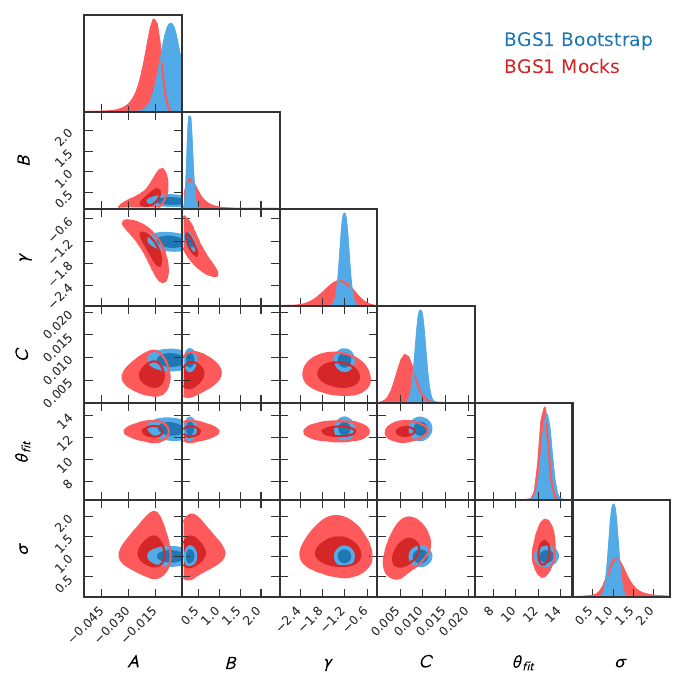}
  \caption{BGS1}\label{fig:bgs_mcmc}
\end{subfigure}\hfill
\begin{subfigure}[t]{0.48\textwidth}
  \centering
  \includegraphics[width=\linewidth]{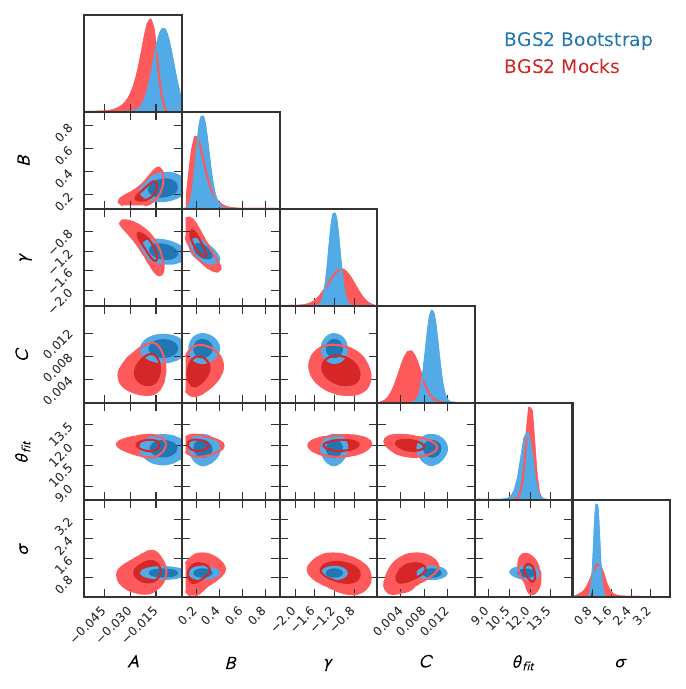}
  \caption{BGS2}\label{fig:bgs_mcmc2}
\end{subfigure}
\caption{BGS Best-fit to the model-free polynomial. Darker: 68\% CL; lighter: 95.45\% CL (MCMC posterior regions).}
\label{fig:combined_mcmc}
\end{figure*}

The parameter $A$ is associated with how the function behaves following the BAO peak. The parameters $B$ and $C$ serve to quantify the significance of their corresponding terms; notably, if there is an absence of a BAO peak, then $C$ will be set to zero. The variable $\gamma$ describes the power-law nature of the function's general form. The physically significant parameters include $\theta_{fit}$ and $\sigma$; specifically, $\theta_{fit}$ determines the location of the BAO, whereas $\sigma$ defines the BAO's width. 
This parametrization has a well-established lineage:  tested by \cite{sanchez2014}, it was first introduced for angular BAO analyses by \cite{carvalho2016baryon}, and applied to low-z samples by \cite{alcaniz2016measuringbaryonacousticoscillations}. Subsequent works \cite{de2018angular,carvalho2020transverse,de2021bao,de2020baryon,menote2022baryon} continued to employ the same functional form across different tracers and redshift ranges, while \cite{marra2019first} adapted it specifically to the radial BAO measurement in redshift space. Within this sequence, the results of \cite{Ferreira2024} stand out as a critical refinement: by analysing the angular correlation function of LRG from the BOSS and eBOSS surveys and incorporating sample covariance directly into the fitting procedure, \cite{Ferreira2024} provide a significantly improved constraint on $\theta_{fit}$
and $\sigma$, effectively isolating the BAO peak with enhanced statistical reliability. Their work demonstrates that the polynomial baseline, particularly the interplay between $C$ and $\gamma$ is essential for marginalizing over systematics that otherwise contaminate the acoustic scale.

We applied flat priors to the parameter $C>0$, $10<\theta_{fit}<20$, $0<\sigma<4$, and $B>0$.
The estimation of parameters was conducted employing the method of maximum likelihood estimation, using the Affine Invariant Markov Chain Monte Carlo (MCMC) Ensemble sampler as implemented in the \texttt{emcee} software, as detailed by \cite{Foreman_Mackey_2013}.  

The sampling strategy employs a hybrid approach for  generation. Approximately 80\% of the proposed moves utilize a Differential Evolution Markov Chain Monte Carlo (DE-MCMC) algorithm \cite{nelson2013run}, implemented in \texttt{emcee} as the \texttt{DEMove}. This method generates a candidate position for a given walker by taking a vector difference between two randomly selected companion walkers from the current ensemble, scaling it, and adding it to a third. The remaining 20\% of moves are governed by the \texttt{DESnooker} algorithm \cite{ter2008differential}. This move is designed to promote global exploration by proposing longer-range jumps. It operates by projecting a walker's position along a line defined by other points in the ensemble. The combination of these two moves leverages the efficiency of the \texttt{DEMove} for detailed sampling while using the \texttt{DESnooker} move to safeguard against poor mixing and premature convergence.

%%----------------------------------------------

\begin{figure*}[H]
\centering
\begin{subfigure}[t]{0.48\textwidth}
  \centering
  \includegraphics[width=\linewidth]{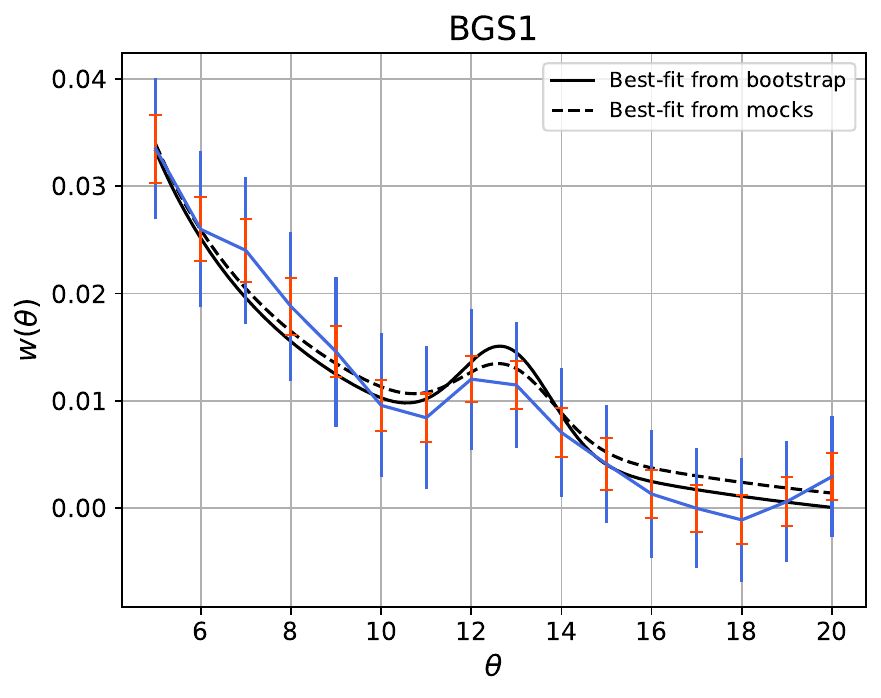}
\end{subfigure}\hfill
\begin{subfigure}[t]{0.48\textwidth}
  \centering
  \includegraphics[width=\linewidth]{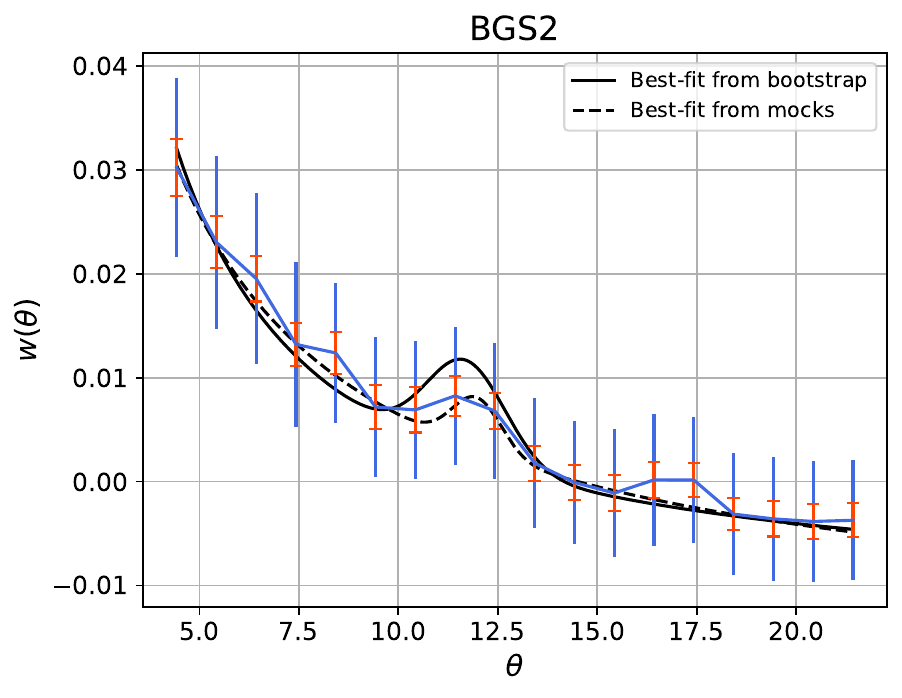}
\end{subfigure}
\caption{Two-point correlation function estimated from the data (blue) and best-fit model (black) using the model-free polynomial fit \citet{sanchez2011tracing}.}
\label{fig:best-fit}
\end{figure*}
%%----------------------------------------------

\section{Results}
We now show the MCMC results for the BGS samples in Figs.~\ref{fig:bgs_mcmc} and~\ref{fig:bgs_mcmc2}, where the red curves correspond to the best-fit obtained using the mock covariance, and the blue curves to those derived from the bootstrap covariance. These results demonstrate that both covariance estimates yield consistent parameter constraints, providing clear evidence of BAO detection, whether the uncertainties are estimated from simulations or from the actual data. For both samples, the \(68\%\) confidence contours exhibit well-defined ellipsoidal shapes for all parameters, indicating that the parameter inference is successful and the likelihood is well behaved. The physical constraints are summarised in Table~\ref{tab:best_fit_results}. In all cases, we obtain \(C>0\) at a significant level, which provides strong evidence for the presence of the BAO peak in the angular correlation function. The reduced \(\chi^2\) values (\(\chi^2/\mathrm{dof}\)) for the mock-based fits are \(2.17\) and \(4.31\) for BGS1 and BGS2, respectively; these values are greater than unity but below 5.0, suggesting that the model provides an adequate description of the data while accounting for the covariance structure. The bootstrap-based fits yield \(\chi^2/\mathrm{dof} = 0.91\) and \(0.68\), which are slightly below unity, indicating conservative error bars.

The best-fit correlation functions are presented in Fig.~\ref{fig:best-fit}. The solid black curve indicates the best-fit model obtained using the bootstrap covariance, while the black dashed curve shows the result based on the mock covariance. The blue curves depict the empirical measurements; the blue error bars are derived from the bootstrap covariance, and the red error bars from the mock covariance. The agreement between the model and the data is excellent for both BGS1 and BGS2, with the best-fit curve accurately tracing the BAO feature and the overall shape of the correlation function. The consistency between the two samples further reinforces the robustness of our results.

\begin{table}[htbp]
\centering
\caption{Best-fit parameter constraints for the BGS samples using mock- and bootstrap-based covariance matrices.}
\begin{tabular}{l c c c c}
\toprule
Parameter & BGS1 (Mocks) & BGS2 (Mocks) & BGS1 (Bootstrap) & BGS2 (Bootstrap) \\
\midrule
\(C\) & \(0.006 \pm 0.002\) & \(0.0056 \pm 0.002\) & \(0.095 \pm 0.009\) & \(0.090 \pm 0.001\) \\
\(\theta_{\rm fit}\) [deg] & \(12.54 \pm 0.23\) & \(11.97 \pm 0.27\) & \(12.77 \pm 0.35\) & \(11.70 \pm 0.42\) \\
\(\sigma\) [deg] & \(1.45 \pm 0.27\) & \(1.03 \pm 0.31\) & \(1.009 \pm 0.094\) & \(0.996 \pm 0.097\) \\
\midrule
\(\chi^2/\mathrm{dof}\) & \(2.17\) & \(4.31\) & \(0.91\) & \(0.68\) \\
\bottomrule
\end{tabular}
\label{tab:best_fit_results}
\end{table}

Several key features emerge from Table~\ref{tab:best_fit_results}. First, the \(\theta_{\rm fit}\) values obtained from the two covariance methods are mutually consistent within the uncertainties, with BGS1 yielding \(\theta_{\rm fit} \approx 12.5^\circ\) and BGS2 yielding \(\theta_{\rm fit} \approx 11.8^\circ\)--\(12.0^\circ\). This consistency confirms that the BAO scale is stable across the two samples and independent of the covariance estimation method. Second, the width parameter \(\sigma\) is better constrained by the bootstrap method, with relative uncertainties of approximately \(10\%\) compared with \(20\%\)--\(30\%\) for the mock-based estimates. This improvement highlights the advantage of the bootstrap approach, which more faithfully captures the actual distribution of the data. Third, the amplitude \(C\) differs significantly between the two methods: the mock-based fits give \(C \sim 0.006\), while the bootstrap fits give \(C \sim 0.09\). This discrepancy arises because the mock-based covariance includes contributions from the fiducial cosmology, whereas the bootstrap covariance is derived entirely from the data, resulting in a different normalisation of the BAO peak amplitude. Nevertheless, the positive value of \(C\) in all cases confirms the detection of the BAO feature.

Overall, the results demonstrate that both BGS samples contain a statistically significant BAO signal, with consistent angular scales and widths. The bootstrap method, being model-independent and data-driven, provides more conservative and reliable error estimates, and we adopt these for our final cosmological analysis.

\subsection{Cosmological inference}

The angular BAO size, $\theta_{\rm BAO}=\frac{r_s}{(1+z) D_A(z)}$, is equal to the known physical size of the sound wave from the early universe ($r_s$) divided by the angular diameter distance to that feature ($D_A(z)$), with an extra factor of $(1+z)$ to account for the expansion of the universe since the light was emitted. For us, $z$ is the effective redshift of the sample, the expected $\theta_{\rm BAO}$ for BGS1 and BGS2 are 9.61 degrees and 8.17 degrees, respectively considering the $\Lambda$CDM model with Planck 18 parameters. For a measured feature, we must correct the BAO bump we found from any projection effect, i.e. correct the $\theta_{fit} $, here we use a model-dependent version and a statistical version, model-independent, for such correction.

For the model dependent version, we apply the correction from \cite{carvalho2016baryon} using \texttt{pyccl} \citep{Chisari_2019}. This involves a shift between $\theta_E(z,\delta z)$, the expected $\theta_{\rm BAO}$ given the considered model, for a tracer at wider bin separation $\delta z$ and the ideal thin-bin case, $\theta_E(z,\delta z =0.002)$. The correction is a normalised difference of the BAO features for the two separations:
\begin{equation}\label{eq:theta_bao}
    \theta_{\rm BAO} = \theta_{fit}(z)+ \alpha(z,\delta z) \theta_{fit}(z)
\end{equation}
where $\alpha$ is
\begin{equation}
    \alpha = \frac{\theta_E(z,\delta z = 0.002)-\theta_E(\delta z)}{\theta_E(z,\delta z = 0.002)}
\end{equation}
%\begin{figure*}
%    \centering
%    \includegraphics[width=\textwidth]{figs/theta_BAO.pdf}
%    \caption{$\theta_{\rm BAO}$ of the three samples of choice.}
%    \label{fig:thetabao}
%\end{figure*}

Due to the 3D measurements from DESI favouring the CPL model, we adjusted our fiducial cosmology choice to determine $\theta_E(z,\delta z = 0.002)$ and examined any potential conflict with \citep{sanchez2011tracing} and the fiducial correlation function. This concern arises because \cite{sanchez2011tracing} indicates a larger discrepancy in the polynomial fit with the exact $\theta_{\rm BAO}$ for $z_{\text{eff}} \leq 0.3$, relevant to BGS1 and BGS2 who have $z_{\text{eff}}\sim 0.2$.

Although this is the usual approach, \cite{Ferreira2024} suggested a purely statistical correction to the projection effects, which also contributes to a more robust and complete model agnostic pipeline. The authors wrote $\alpha$ as a normalised difference between the sample's correlation function and thin bins. The correction parameter is now $\tilde{\alpha}$:
\begin{equation}\label{eq:alpha}
    \tilde{\alpha}_{p}(\theta) = \frac{\displaystyle \sum_{i}^{N_{p}} \mathrm{w}_i(\theta)-\mathrm{w}_{\delta z = 0}(\theta)}{N_{p}},
\end{equation}
where $p$ denotes the percentile of that difference which in this case we chose the median (p=50) and $\tilde{\alpha}_p$ is normalised by ${\sqrt{\int |\mathrm{w}_i(\theta)-\mathrm{w}_{\delta z = 0}(\theta)|^2 \mathrm{d} \theta}}$. considering this correction, we rename $\theta_{\rm BAO}$ as
\begin{equation}\label{eq:theta_bao2}
    \tilde{\theta}_{BAO}(z_{\text{eff}})= \theta_{fit}+ \tilde{\alpha}(\theta_{fit})\theta_{fit}.
\end{equation}

\begin{figure}
    \centering
    \includegraphics[width=\linewidth]{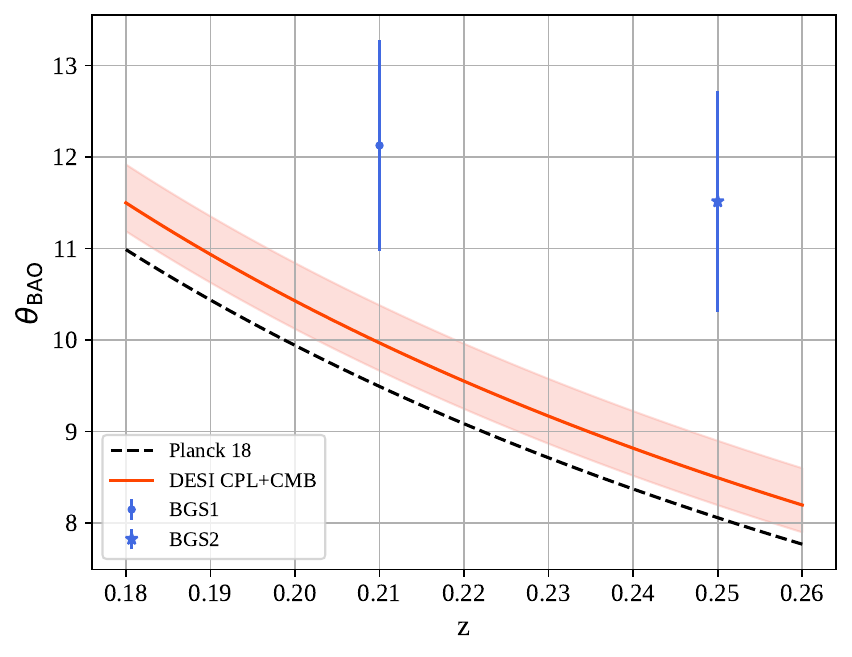}
    \caption{$\theta_{\rm BAO}$ with Planck 18 as the dashed-line and the red line is DESI CPL ($\Omega_m=0.3069\pm0.005$, $H_0=67.97 \pm 0.38$, $w_0=-1.79 \pm ^{0.34}_{0.32}$,$w_a=-1.79\pm^{0.48}_{1.0}$). The observed points are in blue.}
    \label{fig:bao_models}
\end{figure}

The result for $\tilde{\theta}_{BAO}(z_{\text{eff}})$ is shown in Table~\ref{tab:theta_bao_results_final}, we see that the error bars are significantly bigger, this is due to the nature of the method, where we include the actual problems of angular counting with this particular sample and its footprint characteristics, being a more realistic estimation of the uncertainties. Here, using $\tilde{\alpha}$, BGS1 (BGS2) $\tilde{\theta}_{BAO}$ exhibits a 2.84$\sigma$ (3.02$\sigma$) deviation from the expected by Planck 18 $\Lambda$CDM model.
Compared to DESI+CMB constrained CPL parameters \citep{Adame_2025} ($\Omega_m=0.3069\pm0.005$, $H_0=67.97 \pm 0.38$, $w_0=-1.79 \pm ^{0.34}_{0.32}$, $w_a=-1.79\pm 1.0$), the tension for BGS1 $\tilde{\theta}_{BAO}$ is at 2.4$\sigma$ level and BGS2 $\tilde{\theta}_{BAO}$ at 2.7$\sigma$ level, indicating a moderate preference to CPL, but yet not strong. Our result compared to both models can be visualised in Fig.~\ref{fig:bao_models}, in red is the DESI CPL and the dashed line in black is Planck 18, our data points are in blue and the red region is the 68\% CL of DESI results. 
This result is robust considering we are including the imperfections of a not yet complete homogeneously observed footprint, not to mention the problems of using a low redshift sample such as BGS. Again, we must state that a more complete sample in future releases may provide strong indicators to which cosmological model is the best, but for now we understand that the DESI DR1 constraints, using ${\rm w}(\theta)$ as the BAO probe, are inconclusive.
\begin{table}
    \centering
    \begin{tabular}{cc}
    \toprule
         Sample & $\tilde{\theta}_{BAO}(z_{\text{eff}})$ \\
         \hline
         BGS1 & $12.77\pm 1.15$ deg\\
         BGS2 & $11.7\pm 1.21$ deg \\
    \bottomrule
    \end{tabular}
    \caption{BAO result using the statistical projection effect correction from equation \ref{eq:theta_bao2}.}
    \label{tab:theta_bao_results_final}
\end{table}

\section{Summary}

In this work, we studied the angular BAO signature using the DESI BGS samples from the Data Release 1 (DR1). Given that DESI DR1 does not yet provide full sky coverage--a limitation that will be progressively overcome as the survey continues--we selected the optimal samples based on their angular number density on the sky. As a consequence of this careful selection process, only the BGS tracers yielded sufficiently dense and homogeneous samples for a robust angular clustering analysis.

To estimate the covariance matrix, we employed two complementary approaches. First, we constructed 1,000 LPT1 mock catalogues incorporating systematic weights, which served to validate the methodology and provide a cross-check on our error estimates. Although this number of realisations is modest by some standards, it is sufficient for our primary goal of covariance estimation. Second, and more importantly, we adopted a bootstrap resampling procedure as our main method for covariance estimation. For each of 100 bootstrap realisations, we randomly selected 80\% of the survey footprint with replacement, preserving the sky fraction and redshift distribution. The same resampling was applied to the random catalogues used in the correlation function estimator. The bootstrap approach has two decisive advantages: it is entirely model-independent, relying solely on the observed data, and it naturally incorporates the actual survey geometry, fibre-collision patterns, and any residual systematics that may not be fully captured by idealised simulations. For the random catalogues, we developed a novel approach that, by construction, distributes objects isotropically over the observed footprint while being numerically very efficient, as it relies solely on integer-number sampling.

The MCMC analysis was performed using broad and uniform priors on all parameters, thereby avoiding any preferential bias toward specific regions of the \(\theta_{\rm fit}\) parameter space. We adopted the phenomenological model proposed by \citet{sanchez2011tracing} to fit the measured angular correlation function. The resulting reduced \(\chi^2\) values, ranging from \(2\) to \(4\), indicate that the model provides a robust and adequate description of the data. The derived angular BAO scales from the mock-based covariance are \(\theta_{\rm fit} = 12.54^\circ \pm 0.23^\circ\) for BGS1 and \(11.97^\circ \pm 0.27^\circ\) for BGS2, both of which are in significant tension with the expected scale inferred from the Planck 18 \(\Lambda\)CDM cosmology.

However, as noted above, our main results are obtained using the bootstrap covariance matrix, which we consider more robust and data-driven. These bootstrap-based constraints yield \(\theta_{\rm fit} = 12.77^\circ \pm 0.35^\circ\) for BGS1 and \(11.70^\circ \pm 0.42^\circ\) for BGS2. We adopt these as our primary measurements.

We then applied a projection effect correction using cosmological models tested within the DESI framework. This correction, implemented for both \(\Lambda\)CDM and CPL parametrisations, yielded minimal differences in \(\theta_{\rm BAO}\) between the two models. However, this method remains dependent on the assumed cosmological parameters and does not account for the imperfections of the survey footprint. To address these limitations, we adopted the more robust approach of \citet{Ferreira2024}, which performs a purely statistical comparison using thin redshift bins within the sample itself. This method yielded less dramatic tensions with the Planck 18 \(\Lambda\)CDM prediction: \(\theta_{\rm BAO} = 12.77^\circ \pm 1.15^\circ\) for BGS1 and \(11.7^\circ \pm 1.21^\circ\) for BGS2. Interestingly, when compared with the CPL parametrisation, the tensions are reduced to \(2.4\sigma\) for BGS1 and \(2.7\sigma\) for BGS2, which we adopt as our final main result. These findings remain stable even when accounting for the imperfections of an incomplete footprint and the inherent challenges of working with a low-redshift sample such as BGS, where local non-linearities and cosmic variance are more pronounced.

It is worth emphasising that future DESI data releases, with significantly larger sky coverage and improved statistics, will be essential to consolidate these results. With a more complete survey, the angular BAO method presented here will provide increasingly stringent constraints on cosmological parameters and may potentially favour alternative cosmological models. At present, however, we acknowledge that the DESI DR1, using \(w(\theta)\) as the BAO probe, does not yet provide conclusive results. Nevertheless, our work demonstrates the feasibility of extracting robust angular BAO measurements from the BGS sample using both mock-based and bootstrap covariance estimation, and establishes a methodology that can be readily applied to future DESI data releases.
\appendix

\section{Bootstrap spread}

\begin{figure*}[ht]
\centering
\begin{subfigure}[t]{0.48\textwidth}
  \centering
  \includegraphics[width=\linewidth]{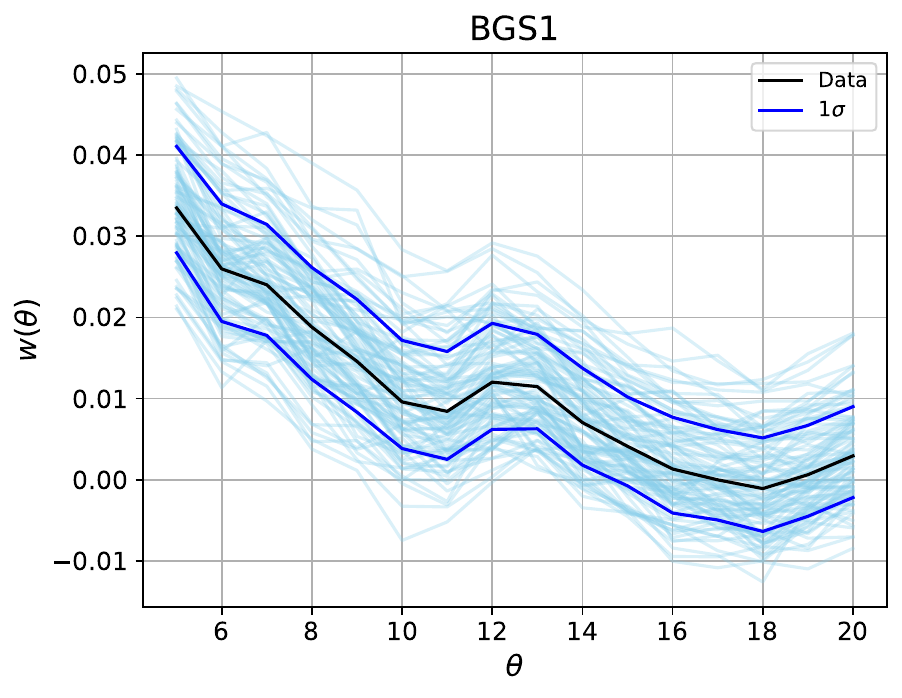}
  \caption{BGS1}\label{fig:bgs1_spread}
\end{subfigure}\hfill
\begin{subfigure}[t]{0.48\textwidth}
  \centering
  \includegraphics[width=\linewidth]{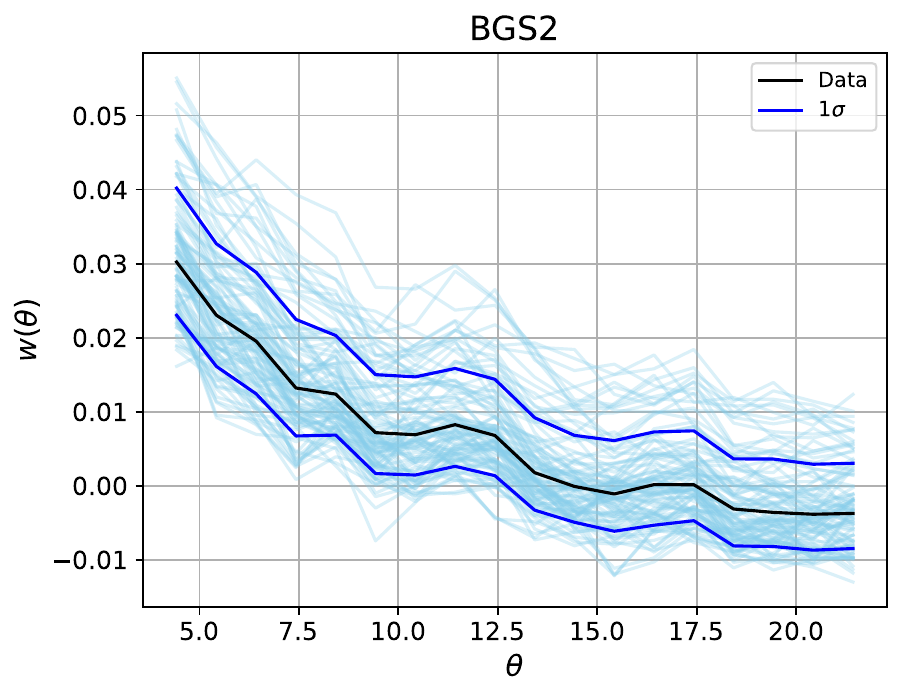}
  \caption{BGS2}\label{fig:bgs2_spread}
\end{subfigure}
\caption{Spread of the angular correlation function \(w(\theta)\) measured from all 100 bootstrap realisations for the BGS1 (top) and BGS2 (bottom) samples. Each light blue line corresponds to one bootstrap resampling, while the black dashed line represents the full bin estimation. The dispersion illustrates the statistical uncertainty captured by the bootstrap covariance matrix. The dark blue curves indicate the 1$\sigma$ confidence level derived from the realizations.}
\label{fig:spread}
\end{figure*}

In this appendix, we present the full set of bootstrap realisations used to estimate the covariance matrix for the BGS samples. Figure~\ref{fig:spread} shows the angular correlation function \(w(\theta)\) measured from each of the 100 bootstrap realisations for BGS1 (top) and BGS2 (bottom). Each grey line represents one bootstrap resampling of the data, while the black dashed line corresponds to the mean correlation function averaged over all realisations.

The spread of the bootstrap realisations provides a visual representation of the statistical uncertainty in our measurement. The dispersion around the mean is larger at small angular scales (\(\theta \lesssim 8^\circ\)), where shot noise and fibre-collision effects are more pronounced, and gradually decreases at larger scales. Importantly, the scatter remains relatively uniform across the angular range of interest (\(8^\circ \lesssim \theta \lesssim 16^\circ\)), indicating that the bootstrap procedure yields a stable estimate of the covariance matrix. The BAO feature, visible as a broad peak around \(\theta \sim 10^\circ\)--\(13^\circ\), is clearly present in the majority of realisations, confirming that the signal is not driven by a particular subsample of the data.

The covariance matrix derived from these bootstrap realisations, as defined in Eq.~\eqref{eq:covbootstrap}, is adopted as our main error estimator throughout this work. As discussed in Section~\ref{sec:methods}, the bootstrap approach has the advantage of being model-independent and naturally incorporating the survey geometry and systematics. The spread shown in Fig.~\ref{fig:spread} confirms that the bootstrap realisations effectively sample the underlying distribution of the data, providing a robust and data-driven estimate of the uncertainties. The consistency between the mean of the bootstrap realisations and the measured correlation function from the full sample (shown in Fig.~\ref{fig:best-fit}) further validates our methodology.

It is worth noting that the amplitude of the BAO feature in individual bootstrap realisations varies within the expected statistical fluctuations. The fact that the BAO peak remains visible in most realisations, even with the relatively small number of objects in the BGS samples, reinforces the statistical significance of our detection. This visual inspection complements the quantitative analysis presented in Table~\ref{tab:best_fit_results}, where the positive values of \(C\) across all realisations confirm the presence of the BAO signal.

\section{Robustness of the mocks}\label{ap:robust_mock}

We ran our MCMC pipeline on the stacked average \(\bar{\omega}(\theta)\) using the same covariance matrix derived from the mock ensemble. The model parameters were allowed to vary freely within the priors defined in Section~\ref{sec:model}. The test was performed for the two main samples analyzed in this work, BGS1 and BGS2.

The recovered \(\theta_{\rm BAO}\) values are compared to the fiducial input values from the mock cosmology in Table~\ref{tab:stacked_recovery}. The uncertainties quoted are the \(1\sigma\) credible intervals from the MCMC posterior on the stacked average. We quantify any potential bias by the deviation in units of sigma, defined as \((\theta_{\rm rec} - \theta_{\rm fid}) / \sigma_{\rm rec}\).

\begin{table}[h]
\centering
\caption{Recovery test on the stacked average mock correlation function. The uncertainties are the \(1\sigma\) posterior credible intervals from the MCMC fit to the stacked data. The deviation is computed as \((\theta_{\rm rec} - \theta_{\rm fid}) / \sigma_{\rm rec}\).}
\begin{tabular}{lcccc}
\hline
Sample & Fiducial \(\theta_{\rm BAO}\) & Recovered \(\theta_{\rm BAO}\) & Deviation (\(\sigma\)) \\
\hline
BGS1 & \(9.67^\circ\) & \(7.80^\circ \pm 2.27^\circ\) & \(0.82\sigma\) \\
BGS2 & \(8.17^\circ\) & \(8.13^\circ \pm 2.22^\circ\) & \(0.02\sigma\) \\
\hline
\end{tabular}
\label{tab:stacked_recovery}
\end{table}

For BGS1, we observe a deviation of \(0.82\sigma\) (\(7.80^\circ\) vs.~\(9.67^\circ\)). For the BGS2 sample, the recovered value \(8.13^\circ \pm 2.22^\circ\) is in excellent agreement with the fiducial input \(8.17^\circ\), with a deviation of only \(0.02\sigma\). This demonstrates that our fitting template does not introduce a significant systematic shift in the BAO scale for these samples with both results compatible in 1$\sigma$ with the expected BAO measurement from the $\Lambda$CDM model used to construct the mocks.

\section{Failed Samples}\label{ap:discard}

In this appendix, we document the samples that were tested but ultimately discarded from our main analysis. While the BGS sample yielded the most reliable angular correlation functions, other DESI tracers - namely LRGs, ELGs, and QSOs - did not meet the necessary criteria for a robust angular BAO detection with the current DESI DR1 data. Additionally, certain BGS redshift bins were excluded due to inconsistencies between the Northern and Southern Galactic Cap (NGC and SGC) measurements.

\subsection{BGS Redshift Bins}

\begin{figure}[htbp]
    \centering
    \includegraphics[width=\linewidth]{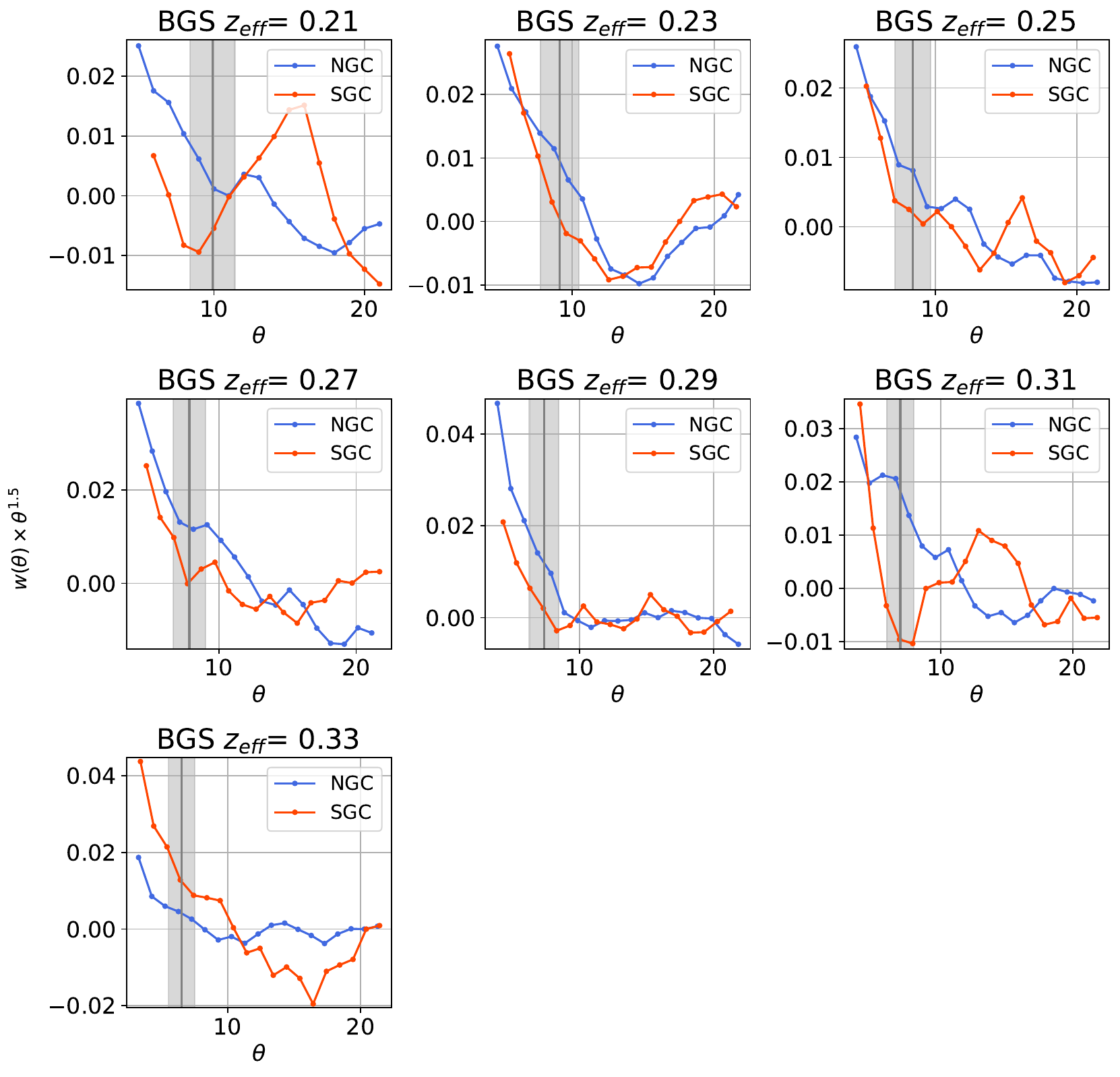}
    \caption{Angular correlation function \(w(\theta)\) for BGS redshift bins from \(z_{\rm eff}=0.21\) to \(0.33\). Solid lines represent NGC measurements, while dashed lines correspond to SGC. The gray vertical lines mark the expected BAO scale derived from the DESI DR1 cosmology, and the shaded gray areas represent the 3$\sigma$ CL. The lowest redshift bins (z \(0.21\)-\(0.25\)) show significant discrepancies between NGC and SGC. The bins at \(z_{\rm eff}=0.31\) and \(0.33\) exhibit unphysical behaviour at small angular scales. Only the NGC1 (\(z_{\rm eff}=0.21\)) and NGC2 (\(z_{\rm eff}=0.25\)) samples were retained for the final analysis.}
    \label{fig:corrfuncs_bgs_ngs_sgc}
\end{figure}

Figure~\ref{fig:corrfuncs_bgs_ngs_sgc} shows the angular correlation function \(w(\theta)\) measured for BGS in several redshift bins, with NGC and SGC measurements displayed separately (blue and red lines, respectively). We included the gray vertical lines mark the expected BAO scale derived from the DESI DR1 cosmology \citep{desicollaboration2025desidr2resultsii}, and the shaded gray areas represent the 3$\sigma$ confidence level (CL). While the BGS sample as a whole provides a robust BAO detection, we found that the lowest redshift bins (\(z_{\rm eff} \lesssim 0.25\)) exhibit significant discrepancies between the NGC and SGC measurements. These discrepancies are likely attributable to the incomplete sky coverage of DESI DR1 and the associated systematics that affect low-redshift clustering measurements. Additionally, the bins at \(z_{\rm eff} = 0.31\) and \(0.33\) show poor agreement with the BAO model, with the correlation function exhibiting an unphysical behaviour at small angular scales. After a systematic evaluation, we selected only the BGS NGC1 and NGC2 samples (at \(z_{\rm eff} = 0.21\) and \(0.25\), respectively) for the final analysis, as these provided the most consistent and well-behaved measurements. Even though the SGC of bins 0.27, 0.29 seems to show a peak near the feature, we dropped from the analysis because the sample had less than 100,000 galaxies  in a smaller sky region, a not consistent sample to represent a fair sample of the sky, as indicated in Table~\ref{tab:data_bins}. 

The separation between the NGC and SGC measurements in the discarded bins, coupled with the absence of the expected BAO signature, underscores our conservative selection. To qualify as a reliable measurement, the angular correlation function must exhibit a characteristic power-law decay at small angular scales, followed by a distinct rise (or bump) at the predicted BAO scale—this inflection clearly signals the acoustic feature above the underlying clustered structure. In the selected NGC bins at \(z_{\rm eff} = 0.21\) and \(0.25\), this decay-and-rise behaviour is consistently observed across the full angular range, whereas the excluded bins either show strong NGC–SGC mismatches or a flattened, unphysical trend that obscures the BAO peak. Thus, we retain only the samples that unambiguously pass this morphological criterion for the final cosmological constraints.

% ------------------------------------------------------------------
\subsubsection{Effective Sky Area and Comoving Volume}
\label{sec:bgs_area_volume}

The NGC and SGC footprints of the BGS sample do not cover the same sky area, and the redshift bins do not sample the same comoving volume. To
quantify this, we compute the effective sky area of each cap and each redshift bin from the union of HEALPix pixels occupied by galaxies in
that sample ($N_{\rm side}=256$), and the comoving volume by scaling the full-sky comoving volume between the bin edges by the fractional sky
area,
\begin{equation}
V_i = \Omega_i \int_{z_{\min}}^{z_{\max}} \frac{dV}{dz\,d\Omega}\,dz ,
\label{eq:comoving_volume_bin}
\end{equation}
where $\Omega_i$ is the effective solid angle of bin $i$.
Table~\ref{tab:bgs_area_volume} reports the results for both caps.

\begin{table*}
\centering
\caption{Effective sky area, comoving volume, and mean galaxy number density for the seven BGS redshift bins, separately for the NGC and SGC.
The area is estimated from the union of \texttt{HEALPix} pixels containing galaxies in each bin ($N_{\rm side}=256$); it is therefore a lower bound
on the true survey footprint. The comoving volume is computed from the fractional sky area using a flat $\Lambda$CDM cosmology with the
Planck~2018 \cite{planck18} parameters. In every redshift bin, the SGC effective area and comoving volume are smaller than the corresponding NGC values by a
factor of $2.4$ to $2.5$, while the mean galaxy number density is only $7$--$13\%$ lower.}
\label{tab:bgs_area_volume}
\begin{tabular}{c c | c c c | c c c}
\hline
 & & \multicolumn{3}{c|}{NGC} & \multicolumn{3}{c}{SGC} \\
Bin & $z$ range
  & Area [$\mathrm{deg}^2$] & Volume [$\mathrm{Gpc}^3$] & $n$ [$\mathrm{Mpc}^{-3}$]
  & Area [$\mathrm{deg}^2$] & Volume [$\mathrm{Gpc}^3$] & $n$ [$\mathrm{Mpc}^{-3}$] \\
\hline
0 & $0.200$--$0.220$ & 4143.12 & 0.07915 & 0.002684 & 1673.45 & 0.03197 & 0.002392 \\
1 & $0.220$--$0.240$ & 4079.65 & 0.09147 & 0.002205 & 1638.62 & 0.03674 & 0.001989 \\
2 & $0.240$--$0.260$ & 3833.11 & 0.09934 & 0.001804 & 1577.77 & 0.04089 & 0.001601 \\
3 & $0.260$--$0.280$ & 3732.08 & 0.11036 & 0.001526 & 1512.88 & 0.04474 & 0.001350 \\
4 & $0.280$--$0.300$ & 3408.06 & 0.11371 & 0.001257 & 1399.47 & 0.04669 & 0.001136 \\
5 & $0.300$--$0.320$ & 3145.31 & 0.11726 & 0.001031 & 1299.38 & 0.04844 & 0.000961 \\
6 & $0.320$--$0.340$ & 2893.78 & 0.11954 & 0.000877 & 1150.09 & 0.04751 & 0.000804 \\
\hline
\end{tabular}
\end{table*}

The NGC sample has an effective area that ranges from $2893.78$ to $4143.12\,\mathrm{deg}^2$ across the seven redshift bins, a factor of
$1.43$. The comoving volume ranges from $0.07915$ to $0.11954\,\mathrm{Gpc}^3$, a factor of $1.51$. The mean galaxy number
density ranges from $0.000877$ to $0.002684\,\mathrm{Mpc}^{-3}$, a factor of $3.06$. The SGC sample has effective areas between $1150.09$ and
$1673.45\,\mathrm{deg}^2$ and comoving volumes between $0.03197$ and $0.04844\,\mathrm{Gpc}^3$. In every redshift bin, the SGC effective area
and comoving volume are smaller than the corresponding NGC values by a factor of $2.4$ to $2.5$, while the mean galaxy number density is only
$7$--$13\%$ lower. This demonstrates that the BGS bins do not sample equal sky areas or equal comoving volumes, and that the two caps do not
sample the same volume at fixed redshift. Since BGS is covering the smallest redshift bins, the BAO are above 5 deg, thus the sky coverage is an important feature to be considered.

The expected BAO signal-to-noise scales approximately as $\sqrt{V_{\rm eff}}\,n_{\rm bar}$, so the SGC bins in this redshift range
have an expected signal-to-noise that is only $\sim58\%$ of the corresponding NGC bins. Equivalently, the NGC sample provides a factor of
$\sim1.7$ more statistical power for BAO detection than the SGC at fixed redshift. This explains why the SGC clustering amplitudes around
$\theta\sim8^\circ$ appear noisier than the NGC ones at the same angular scale, and why no reliable cosmological inference can be drawn from the
SGC bins alone at the current sample size.

% ------------------------------------------------------------------
\subsubsection{Uniform Selection Criterion for the BAO Bins}
\label{sec:bgs_selection}

Selecting redshift bins on the basis of a visible BAO peak is circular and would introduce confirmation bias. We therefore replace the visual
selection with a uniform, quantitative criterion applied to all seven BGS bins before any cosmological inference is drawn. The criterion is
defined entirely in terms of the shape, scale, and significance of the fitted BAO feature, and is applied identically to every bin.

We fit the \citet{sanchez2011tracing} model to the NGC angular correlation function of each bin. The no-BAO model is the same fit with $C=0$. To avoid local minima, each fit is restarted from a grid of initial guesses spanning the full prior range in $\theta_{\rm BAO}$ and
$\sigma$, and the solution with the lowest $\chi^2$ is retained.

A bin is successful when containing a BAO feature only if all four of the following conditions are satisfied:

\begin{enumerate}
  \item The Gaussian amplitude is positive and the improvement in
        $\chi^2$ relative to the no-BAO model satisfies
        $\Delta\chi^2 > 4$.
  \item The fitted width satisfies $0.3^\circ < \sigma < 2.0^\circ$,
        consistent with the expected BAO width in a narrow redshift bin.
  \item The fitted centroid satisfies
        $|\theta_{\rm BAO} - r_{\rm drag}/D_A(z)| \,/\,
        (r_{\rm drag}/D_A(z)) < 0.30$, where $r_{\rm drag}=147.09\,\mathrm{Mpc}$
        and $D_A(z)$ is the Planck~2018 angular diameter distance.
  \item The fitted $\theta_{\rm BAO}$ lies at least $1\sigma$ inside the
        prior range, so that the fit is not saturating at a boundary.
\end{enumerate}

Table~\ref{tab:bgs_sanchez} reports the best-fit parameters for each bin. Only the bins at $z_{\rm eff}=0.21$ and $z_{\rm eff}=0.25$ satisfy
all four criteria. The remaining bins fail for specific, quantifiable reasons. The bin at $z_{\rm eff}=0.23$ has $\theta_{\rm BAO}$ pinned at
the upper prior boundary and $\sigma=2.28^\circ$ above the allowed range. The bin at $z_{\rm eff}=0.27$ has $\sigma=1.25^\circ$, a factor
of $2.5$ broader than the bin at $z_{\rm eff}=0.25$, and its centroid lies within $1\sigma$ of the lower prior boundary. The bin at
$z_{\rm eff}=0.29$ has $\sigma=3.99^\circ$ pinned at the upper boundary. The bin at $z_{\rm eff}=0.31$ has $\theta_{\rm BAO}$ pinned at the upper
boundary. The bin at $z_{\rm eff}=0.33$ has a centroid inconsistent with the expected $r_{\rm drag}/D_A(z)$ at that redshift. The two selected
bins have narrow Gaussian features ($\sigma=0.77^\circ$ and $0.50^\circ$), centroids consistent with the expected
$r_{\rm drag}/D_A(z)$ trend, and no parameter saturation.

\begin{table}
\centering
\caption{\citet{sanchez2011tracing} fit to the NGC angular correlation
function of each BGS redshift bin. Columns: bin index; effective
redshift; best-fit Gaussian amplitude $C$; best-fit centroid
$\theta_{\rm BAO}$; best-fit width $\sigma$; improvement in $\chi^2$
relative to a smooth no-BAO model; and whether the bin satisfies all
four selection criteria defined in the text. Parameters marked with a
dagger are saturated at a prior boundary. Only the bins at
$z_{\rm eff}=0.21$ and $z_{\rm eff}=0.25$ satisfy all criteria.}
\label{tab:bgs_sanchez}
\begin{tabular}{c c c c c c c}
\hline
Bin & $z_{\rm eff}$ & $C$ & $\theta_{\rm BAO}$ [deg] & $\sigma$ [deg] & $\Delta\chi^2$ & Selected \\
\hline
0 & 0.21 & 0.0060 & 12.73 & 0.77 & 7.07 & yes \\
1 & 0.23 & 0.0204 & $20.00^{\dagger}$ & 2.28 & 54.9 & no \\
2 & 0.25 & 0.0057 & 11.93 & 0.50 & 5.65 & yes \\
3 & 0.27 & 0.0067 & $10.06^{\dagger}$ & 1.25 & 5.83 & no \\
4 & 0.29 & 0.0155 & 18.35 & $3.99^{\dagger}$ & 3.15 & no \\
5 & 0.31 & 0.0127 & $20.00^{\dagger}$ & 2.03 & 24.4 & no \\
6 & 0.33 & 0.0032 & 14.10 & 0.82 & 1.94 & no \\
\hline
\end{tabular}
\\[4pt]
{\footnotesize $^{\dagger}$Parameter saturated at a prior boundary.}
\end{table}

\subsection{Luminous Red Galaxies}

\begin{figure}[htbp]
    \centering
    \includegraphics[width=\linewidth]{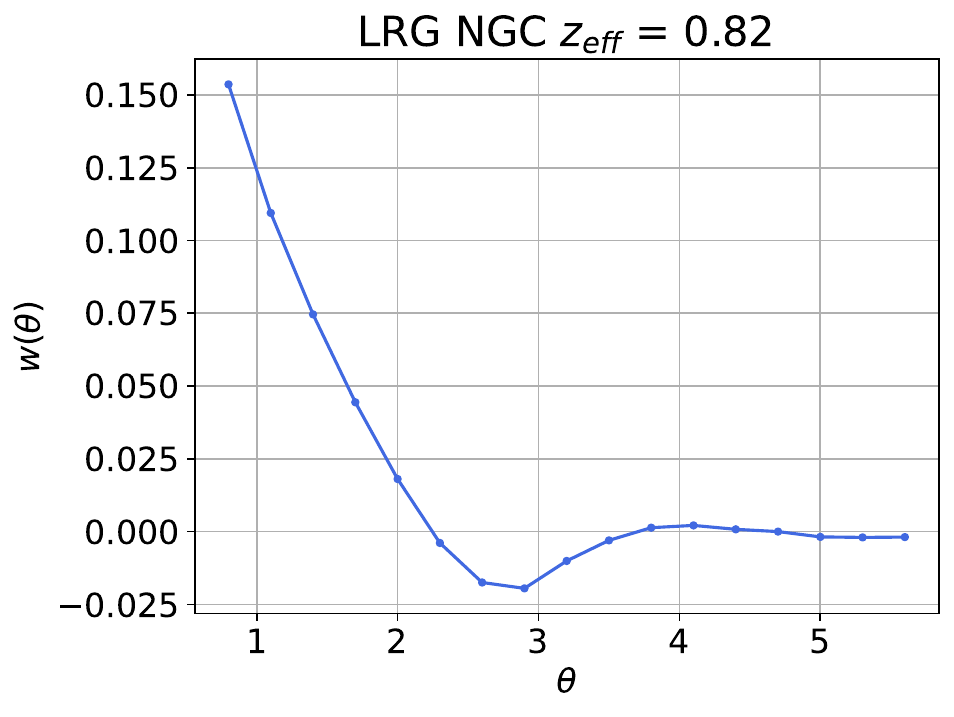}
    \caption{Angular correlation function \(w(\theta)\) for the LRG sample. The signal is dominated by noise and does not exhibit a clear BAO feature. The amplitude of the correlation function is very low, and the error bars are too large for a meaningful detection. The LRG sample was discarded due to insufficient signal-to-noise ratio.}
    \label{fig:corrfuncs_lrg}
\end{figure}

Figure~\ref{fig:corrfuncs_lrg} presents the angular correlation function for the LRG sample. Despite the high spectroscopic completeness of the LRG target selection \citep{Zhou2023}, we found the angular clustering signal to be too weak for a reliable BAO detection with the current DESI DR1 footprint. The correlation function shows no clear BAO feature, and the signal-to-noise ratio is insufficient for a meaningful extraction of \(\theta_{\rm BAO}\). This is likely due to the lower angular number density of LRGs compared to BGS, combined with the limited sky coverage of the early data release. We therefore excluded the LRG sample from our analysis.

\subsection{Quasars}

\begin{figure}[htbp]
    \centering
    \includegraphics[width=\linewidth]{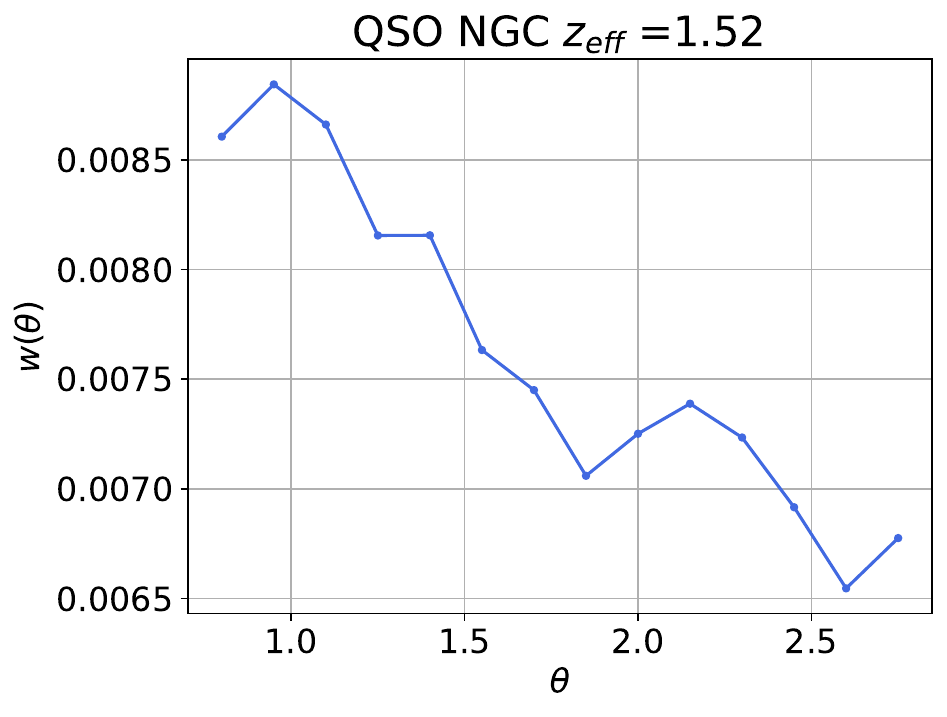}
    \caption{Angular correlation function \(w(\theta)\) for the QSO sample at \(z_{\rm eff}=1.52\). The signal is consistent with zero within the uncertainties, with no BAO feature visible. The low number density of quasars leads to prohibitive shot noise, and projection effects further suppress the signal. The QSO sample was excluded from the analysis.}
    \label{fig:corrfuncs_qso}
\end{figure}

The QSO sample, shown in Fig.~\ref{fig:corrfuncs_qso}, yielded the weakest clustering signal among all tracers. At \(z_{\rm eff} = 1.52\), the angular correlation function is consistent with zero within the uncertainties, with no discernible BAO feature. The low number density of quasars (\(\sim 200\) deg\(^{-2}\)) and the sparse sampling of the high-redshift Universe result in prohibitive shot noise for the angular clustering measurement. Moreover, the projection effects are particularly severe for the QSO sample, as the broad redshift distribution washes out the BAO signal in the angular projection. For these reasons, the QSO sample was excluded from our analysis.

In summary, the sample selection process for our angular BAO analysis was necessarily conservative. Of the four DESI tracers and the multiple redshift bins tested, only the BGS NGC1 and NGC2 samples provided sufficiently robust angular correlation functions for a reliable BAO measurement. While LRGs, QSOs, and the remaining BGS bins may become viable with future DESI data releases - which will offer larger sky coverage and improved statistics - the current DR1 data do not support their use for the angular BAO method presented in this work.

\section*{Acknowledgements}
Paula Ferreira acknowledges the support from the start-up funding of Zhejiang University, Zhejiang provincial top level research support program and the Brazilian funding agency FINEP for the contract under the MultiGPU project 1891/22. Pedro da Silveira Ferreira thanks acknowledges the support from the start-up funding of Zhejiang University, Zhejiang provincial top level research support program and the Brazilian funding agency FINEP (contract 01.22.0505.00) for financial support. This work made use, mostly, of the Milliways HPC computer located at the Instituto de Física in the Universidade Federal do Rio de Janeiro, managed and funded by \href{https://sites.google.com/view/arcos-ufrj/about?authuser=0}{ARCOS} (Astrophysics, Relativity and COSmology research group). The authors made use of one of the Sci-Mind servers machines, developed by the CBPF Lab-IA team. UR thanks the Brazilian National Research Council (CNPq) for the financial support. CRB acknowledges the financial support from CNPq (316072/2021-4) and from FAPERJ (grants 201.456/2022 and 210.330/2022) and the FINEP contract 01.22.0505.00 (ref. 1891/22). AB acknowledges a CNPq fellowship. 

This research used data obtained with the Dark Energy Spectroscopic Instrument (DESI). DESI construction and operations is managed by the Lawrence Berkeley National Laboratory. This material is based upon work supported by the U.S. Department of Energy, Office of Science, Office of High-Energy Physics, under Contract No. DE–AC02–05CH11231, and by the National Energy Research Scientific Computing Center, a DOE Office of Science User Facility under the same contract. Additional support for DESI was provided by the U.S. National Science Foundation (NSF), Division of Astronomical Sciences under Contract No. AST-0950945 to the NSF’s National Optical-Infrared Astronomy Research Laboratory; the Science and Technology Facilities Council of the United Kingdom; the Gordon and Betty Moore Foundation; the Heising-Simons Foundation; the French Alternative Energies and Atomic Energy Commission (CEA); the National Council of Humanities, Science and Technology of Mexico (CONAHCYT); the Ministry of Science and Innovation of Spain (MICINN), and by the DESI Member Institutions: www.desi.lbl.gov/collaborating-institutions. The DESI collaboration is honored to be permitted to conduct scientific research on I’oligam Du’ag (Kitt Peak), a mountain with particular significance to the Tohono O’odham Nation. Any opinions, findings, and conclusions or recommendations expressed in this material are those of the author(s) and do not necessarily reflect the views of the U.S. National Science Foundation, the U.S. Department of Energy, or any of the listed funding agencies.

\printcredits

\bibliographystyle{cas-model2-names}

\bibliography{cas-refs}

@article{ramirez2022colore,
  title={CoLoRe: fast cosmological realisations over large volumes with multiple tracers},
  author={Ram{\'\i}rez-P{\'e}rez, C{\'e}sar and  others},
  journal={Journal of Cosmology and Astroparticle Physics},
  volume={2022},
  number={05},
  pages={002},
  year={2022},
  publisher={IOP Publishing},
  doi={10.1088/1475-7516/2022/05/002}
}

@article{chuang2015ezmocks,
  title={EZmocks: extending the Zel'dovich approximation to generate mock galaxy catalogues with accurate clustering statistics},
  author={Chuang, Chia-Hsun and others},
  journal={Monthly Notices of the Royal Astronomical Society},
  volume={446},
  number={3},
  pages={2621--2628},
  year={2015},
  publisher={Oxford University Press},
doi={10.1093/mnras/stu2301}
}

@article{bernardeau2002large,
  title={Large-scale structure of the Universe and cosmological perturbation theory},
  author={Bernardeau, Francis and others},
  journal = {Physics Reports},
volume = {367},
number = {1},
pages = {1-248},
year = {2002},
issn = {0370-1573},
doi = {10.1016/S0370-1573(02)00135-7},
}

@article{sanchez2011tracing,
  title={Tracing the sound horizon scale with photometric redshift surveys},
  author={S{\'a}nchez, Eusebio and others},
  journal={Monthly Notices of the Royal Astronomical Society},
  volume={411},
  number={1},
  pages={277--288},
  year={2011},
  publisher={Blackwell Publishing Ltd Oxford, UK},
doi={10.1111/j.1365-2966.2010.17679.x}
}

@article{pinon2025mitigation,
  title={Mitigation of DESI fiber assignment incompleteness effect on two-point clustering with small angular scale truncated estimators},
  author={Pinon, M and others},
  journal={Journal of Cosmology and Astroparticle Physics},
  volume={2025},
  number={01},
  pages={131},
  year={2025},
  publisher={IOP Publishing},
doi={10.1088/1475-7516/2025/01/131}
}

@article{landy1993bias,
  title={Bias and variance of angular correlation functions},
  author={Landy, Stephen D and Szalay, Alexander S},
  journal={The Astrophysical Journal},
  volume={412},
  pages={64--71},
  year={1993},
doi={10.1086/172900}
}

@article{eisenstein_hu,
  title={Baryonic features in the matter transfer function},
  author={Eisenstein, Daniel J and Hu, Wayne},
  journal={The Astrophysical Journal},
  volume={496},
  number={2},
  pages={605--614},
  year={1998}
}

@article{peebles1973statistical,
  title={Statistical analysis of catalogs of extragalactic objects. I. Theory},
  author={Peebles, PJE},
  journal={Astrophysical Journal, Vol. 185, pp. 413-440 (1973)},
  volume={185},
  pages={413--440},
  year={1973},
doi={10.1086/152431}
}

@article{novell2025full,
  title={Full-Shape analysis of the power spectrum and bispectrum of DESI DR1 LRG and QSO samples},
  author={Novell-Masot, S and others},
  journal={Journal of Cosmology and Astroparticle Physics},
  volume={2025},
  number={06},
  pages={005},
  year={2025},
  publisher={IOP Publishing},
doi={10.1088/1475-7516/2025/06/005}
}

@article{carvalho2016baryon,
  title={Baryon acoustic oscillations from the SDSS DR10 galaxies angular correlation function},
  author={Carvalho, GC and others},
  journal={Physical Review D},
  volume={93},
  number={2},
  pages={023530},
  year={2016},
  publisher={APS},
doi={0.1103/PhysRevD.93.023530}
}

@misc{alcaniz2016measuringbaryonacousticoscillations,
      title={Measuring baryon acoustic oscillations with angular two-point correlation function}, 
      author={Jailson S. Alcaniz and others},
      year={2016},
      eprint={1611.08458},
      archivePrefix={arXiv},
      primaryClass={astro-ph.CO},
      url={https://arxiv.org/abs/1611.08458}, 
}

@article{de2018angular,
  title={Angular Baryon Acoustic Oscillation measure at z= 2.225 from the SDSS quasar survey},
  author={De Carvalho, E and others},
  journal={Journal of Cosmology and Astroparticle Physics},
  volume={2018},
  number={04},
  pages={064},
  year={2018},
  publisher={IOP Publishing},
 doi={10.1088/1475-7516/2018/04/064}
}

@article{carvalho2020transverse,
  title={The transverse baryonic acoustic scale from the SDSS DR11 galaxies},
  author={Carvalho, GC and others},
  journal={Astroparticle Physics},
  volume={119},
  pages={102432},
  year={2020},
  publisher={Elsevier},
 doi={10.1016/j.astropartphys.2020.102432}
}

@article{de2021bao,
  title={BAO angular scale at zeff= 0.11 with the SDSS blue galaxies},
  author={De Carvalho, E and others},
  journal={Astronomy \& Astrophysics},
  volume={649},
  pages={A20},
  year={2021},
  publisher={EDP Sciences},
doi={10.1051/0004-6361/202039936}
}

@article{de2020baryon,
  title={Baryon acoustic oscillations signature in the three-point angular correlation function from the SDSS-DR12 quasar survey},
  author={De Carvalho, E and others},
  journal={Monthly Notices of the Royal Astronomical Society},
  volume={492},
  number={3},
  pages={4469--4476},
  year={2020},
  publisher={Oxford University Press},
doi={10.1093/mnras/staa119}
}

@article{menote2022baryon,
  title={Baryon acoustic oscillations in thin redshift shells from BOSS DR12 and eBOSS DR16 galaxies},
  author={Menote, Ranier and Marra, Valerio},
  journal={Monthly Notices of the Royal Astronomical Society},
  volume={513},
  number={2},
  pages={1600--1608},
  year={2022},
  publisher={Oxford University Press},
doi={10.1093/mnras/stac847}
}

@article{Ferreira2024,
author={Ferreira, Paula S.
and Reis, Ribamar R. R.},
title={Angular correlation function from sample covariance with BOSS and eBOSS LRG},
journal={The European Physical Journal C},
year={2024},
month={May},
day={06},
volume={84},
number={5},
pages={466},
issn={1434-6052},
doi={10.1140/epjc/s10052-024-12808-4},
}

@article{marra2019first,
  title={A first model-independent radial BAO constraint from the final BOSS sample},
  author={Marra, Valerio and Chirinos Isidro, Eddy G},
  journal={Monthly Notices of the Royal Astronomical Society},
  volume={487},
  number={3},
  pages={3419--3426},
  year={2019},
  publisher={Oxford University Press},
doi={10.1093/mnras/stz1557}
}

@ARTICLE{Eisenstein05,
       author = {Eisenstein, Daniel J. and others},
        title = "{Detection of the Baryon Acoustic Peak in the Large-Scale Correlation Function of SDSS Luminous Red Galaxies}",
      journal = {apj},
         year = 2005,
        month = nov,
       volume = {633},
       number = {2},
        pages = {560-574},
          doi = {10.1086/466512},
archivePrefix = {arXiv},
       eprint = {astro-ph/0501171},
 primaryClass = {astro-ph},
       adsurl = {https://ui.adsabs.harvard.edu/abs/2005ApJ...633..560E}
}

@article{sanchez2014,
    author = {Salazar-Albornoz, Salvador and others},
    title = {Clustering tomography: measuring cosmological distances through angular clustering in thin redshift shells},
    journal = {Monthly Notices of the Royal Astronomical Society},
    volume = {443},
    number = {4},
    pages = {3612-3623},
    year = {2014},
    month = {08},
    doi = {10.1093/mnras/stu1428},
}

@misc{desicollaboration2025datarelease1dark,
      title={Data Release 1 of the Dark Energy Spectroscopic Instrument}, 
      author={{DESI Collaboration}},
      year={2025},
      eprint={2503.14745},
      archivePrefix={arXiv},
      primaryClass={astro-ph.CO},
      url={https://arxiv.org/abs/2503.14745}, 
}

@article{Anonymous_2025,
   title={DESI DR2 results. I. Baryon acoustic oscillations from the Lyman alpha forest},
   ISSN={2470-0029},
   url={http://dx.doi.org/10.1103/2wwn-xjm5},
   DOI={10.1103/2wwn-xjm5},
   journal={Physical Review D},
   publisher={American Physical Society (APS)},
   author={{DESI Collaboration}},
   year={2025},
   month=jun }

@misc{desicollaboration2025desidr2resultsii,
      title={DESI DR2 Results II: Measurements of Baryon Acoustic Oscillations and Cosmological Constraints}, 
      author={{DESI Collaboration}},
      year={2025},
      eprint={2503.14738},
      archivePrefix={arXiv},
      primaryClass={astro-ph.CO},
      url={https://arxiv.org/abs/2503.14738}, 
}

@article{Foreman_Mackey_2013,
   title={emcee: The MCMC Hammer},
   volume={125},
   ISSN={1538-3873},
   url={http://dx.doi.org/10.1086/670067},
   DOI={10.1086/670067},
   number={925},
   journal={Publications of the Astronomical Society of the Pacific},
   publisher={IOP Publishing},
   author={Foreman-Mackey, Daniel and others},
   year={2013},
   month=mar, pages={306–312} }

@article{Hartlap_2006,
   title={Why your model parameter confidences might be too optimistic.  Unbiased estimation of the inverse covariance matrix},
   volume={464},
   ISSN={1432-0746},
   url={http://dx.doi.org/10.1051/0004-6361:20066170},
   DOI={10.1051/0004-6361:20066170},
   number={1},
   journal={Astronomy \& Astrophysics},
   publisher={EDP Sciences},
   author={Hartlap, J. and Simon, P. and Schneider, P.},
   year={2006},
   month=dec, pages={399–404} }

@ARTICLE{Cole05,
       author = {{Cole}, Shaun and others},
        title = "{The 2dF Galaxy Redshift Survey: power-spectrum analysis of the final data set and cosmological implications}",
      journal = {mnras},
         year = 2005,
        month = sep,
       volume = {362},
       number = {2},
        pages = {505-534},
          doi = {10.1111/j.1365-2966.2005.09318.x},
archivePrefix = {arXiv},
       eprint = {astro-ph/0501174},
 primaryClass = {astro-ph},
       adsurl = {https://ui.adsabs.harvard.edu/abs/2005MNRAS.362..505C}
}

@ARTICLE{planck18,
       author = {{Planck Collab.}},
        title = "{Planck 2018 results. I. Overview and the cosmological legacy of Planck}",
      journal = {Astronomy and Astrophysics},
         year = 2020,
        month = sep,
       volume = {641},
          eid = {A1},
        pages = {A1},
          doi = {10.1051/0004-6361/201833880},
archivePrefix = {arXiv},
       eprint = {1807.06205},
 primaryClass = {astro-ph.CO},
       adsurl = {https://ui.adsabs.harvard.edu/abs/2020A&A...641A...1P}
}

@article{Chisari_2019,
   title={Core Cosmology Library: Precision Cosmological Predictions for LSST},
   volume={242},
   ISSN={1538-4365},
   url={http://dx.doi.org/10.3847/1538-4365/ab1658},
   DOI={10.3847/1538-4365/ab1658},
   number={1},
   journal={The Astrophysical Journal Supplement Series},
   publisher={American Astronomical Society},
   author={Chisari, Nora Elisa and others},
   year={2019},
   month=may, pages={2} }

@article{Adame_2025,
doi = {10.1088/1475-7516/2025/02/021},
url = {https://dx.doi.org/10.1088/1475-7516/2025/02/021},
year = {2025},
month = {feb},
publisher = {IOP Publishing},
volume = {2025},
number = {02},
pages = {021},
author = {{DESI collaboration}},
title = {DESI 2024 VI:  cosmological constraints from the measurements of baryon acoustic oscillations},
journal = {Journal of Cosmology and Astroparticle Physics}
}

@misc{levi2019darkenergyspectroscopicinstrument,
      title={The Dark Energy Spectroscopic Instrument (DESI)}, 
      author={Michael E. Levi and others},
      year={2019},
      eprint={1907.10688},
      archivePrefix={arXiv},
      primaryClass={astro-ph.IM},
      url={https://arxiv.org/abs/1907.10688}, 
}

@article{Ross_2025,
doi = {10.1088/1475-7516/2025/01/125},
url = {https://dx.doi.org/10.1088/1475-7516/2025/01/125},
year = {2025},
month = {jan},
publisher = {IOP Publishing},
volume = {2025},
number = {01},
pages = {125},
author = {Ross, A.J. and others},
title = {The construction of large-scale structure catalogs for the Dark Energy Spectroscopic Instrument},
journal = {Journal of Cosmology and Astroparticle Physics},
}

@ARTICLE{fkp94,
       author = {Feldman, Hume A. and Kaiser, Nick and Peacock, John A.},
        title = "{Power-Spectrum Analysis of Three-dimensional Redshift Surveys}",
      journal = {APJ},
         year = 1994,
        month = may,
       volume = {426},
        pages = {23},
          doi = {10.1086/174036},
archivePrefix = {arXiv},
       eprint = {astro-ph/9304022},
 primaryClass = {astro-ph},
       adsurl = {https://ui.adsabs.harvard.edu/abs/1994ApJ...426...23F}
}

@article{nelson2013run,
   title={RUN DMC: AN EFFICIENT, PARALLEL CODE FOR ANALYZING RADIAL VELOCITY OBSERVATIONS USING  N-BODY INTEGRATIONS AND DIFFERENTIAL EVOLUTION MARKOV CHAIN MONTE CARLO},
   volume={210},
   ISSN={1538-4365},
   DOI={10.1088/0067-0049/210/1/11},
   number={1},
   journal={The Astrophysical Journal Supplement Series},
   publisher={American Astronomical Society},
   author={Nelson, Benjamin and Ford, Eric B. and Payne, Matthew J.},
   year={2013}}

@article{ter2008differential,
  title={Differential evolution Markov chain with snooker updater and fewer chains},
  author={Ter Braak, Cajo JF and Vrugt, Jasper A},
  journal={Statistics and Computing},
  doi={10.1007/s11222-008-9104-9},
  volume={18},
  number={4},
  pages={435--446},
  year={2008},
  publisher={Springer}
}

@article{Cole_2005,
   title={The 2dF Galaxy Redshift Survey: power-spectrum analysis of the final data set and cosmological implications},
   volume={362},
   ISSN={1365-2966},
   url={http://dx.doi.org/10.1111/j.1365-2966.2005.09318.x},
   DOI={10.1111/j.1365-2966.2005.09318.x},
   number={2},
   journal={Monthly Notices of the Royal Astronomical Society},
   publisher={Oxford University Press (OUP)},
   author={Cole, Shaun and others},
   year={2005},
   month=sep, pages={505–534} }

@article{Percival_2010,
   title={Baryon acoustic oscillations in the Sloan Digital Sky Survey Data Release 7 galaxy sample},
   volume={401},
   ISSN={1365-2966},
   url={http://dx.doi.org/10.1111/j.1365-2966.2009.15812.x},
   DOI={10.1111/j.1365-2966.2009.15812.x},
   number={4},
   journal={Monthly Notices of the Royal Astronomical Society},
   publisher={Oxford University Press (OUP)},
   author={Percival, Will J. and others},
   year={2010},
   month=feb, pages={2148–2168} }

@article{Alam_2017,
   title={The clustering of galaxies in the completed SDSS-III Baryon Oscillation Spectroscopic Survey: cosmological analysis of the DR12 galaxy sample},
   volume={470},
   ISSN={1365-2966},
   url={http://dx.doi.org/10.1093/mnras/stx721},
   DOI={10.1093/mnras/stx721},
   number={3},
   journal={Monthly Notices of the Royal Astronomical Society},
   publisher={Oxford University Press (OUP)},
   author={Alam, Shadab and others},
   year={2017},
   month=mar, pages={2617–2652} }

@article{Gil_Mar_n_2020,
   title={The Completed SDSS-IV extended Baryon Oscillation Spectroscopic Survey: measurement of the BAO and growth rate of structure of the luminous red galaxy sample from the anisotropic power spectrum between redshifts 0.6 and 1.0},
   volume={498},
   ISSN={1365-2966},
   url={http://dx.doi.org/10.1093/mnras/staa2455},
   DOI={10.1093/mnras/staa2455},
   number={2},
   journal={Monthly Notices of the Royal Astronomical Society},
   publisher={Oxford University Press (OUP)},
   author={Gil-Marín, Héctor and others},
   year={2020},
   month=aug, pages={2492–2531} }

@article{Chan_2022,
   title={Dark Energy Survey Year 3 results: Measurement of the baryon acoustic oscillations with three-dimensional clustering},
   volume={106},
   ISSN={2470-0029},
   url={http://dx.doi.org/10.1103/PhysRevD.106.123502},
   DOI={10.1103/physrevd.106.123502},
   number={12},
   journal={Physical Review D},
   publisher={American Physical Society (APS)},
   author={Chan, K. C. and others},
   year={2022},
   month=dec }

@article{CHEVALLIER_2001,
   title={ACCELERATING UNIVERSES WITH SCALING DARK MATTER},
   volume={10},
   ISSN={1793-6594},
   url={http://dx.doi.org/10.1142/S0218271801000822},
   DOI={10.1142/s0218271801000822},
   number={02},
   journal={International Journal of Modern Physics D},
   publisher={World Scientific Pub Co Pte Lt},
   author={Chevallier, Michel and Polarski, David},
   year={2001},
   month=apr, pages={213–223} }

@article{Linder_2003,
   title={Exploring the Expansion History of the Universe},
   volume={90},
   ISSN={1079-7114},
   url={http://dx.doi.org/10.1103/PhysRevLett.90.091301},
   DOI={10.1103/physrevlett.90.091301},
   number={9},
   journal={Physical Review Letters},
   publisher={American Physical Society (APS)},
   author={Linder, Eric V.},
   year={2003},
   month=mar }

@article{Novell-Masot_2025,
doi = {10.1088/1475-7516/2025/06/005},
url = {https://dx.doi.org/10.1088/1475-7516/2025/06/005},
year = {2025},
month = {jun},
publisher = {IOP Publishing},
volume = {2025},
number = {06},
pages = {005},
author = {Novell-Masot, S. and others},
title = {Full-Shape analysis of the power spectrum and bispectrum of DESI DR1 LRG and QSO samples},
journal = {Journal of Cosmology and Astroparticle Physics},

}

@article{Di_Valentino_2025,
   title={The CosmoVerse White Paper: Addressing observational tensions in cosmology with systematics and fundamental physics},
   volume={49},
   ISSN={2212-6864},
   url={http://dx.doi.org/10.1016/j.dark.2025.101965},
   DOI={10.1016/j.dark.2025.101965},
   journal={Physics of the Dark Universe},
   publisher={Elsevier BV},
   author={Di Valentino, Eleonora and others},
   year={2025},
   month=sep, pages={101965} }

@ARTICLE{Hahn2023,
       author = {{Hahn}, ChangHoon and others},
        title = "{The DESI Bright Galaxy Survey: Final Target Selection, Design, and Validation}",
      journal = {The Astronomical Journal},
         year = 2023,
        month = jun,
       volume = {165},
       number = {6},
          eid = {253},
        pages = {253},
          doi = {10.3847/1538-3881/accff8},
archivePrefix = {arXiv},
       eprint = {2208.08512},
 primaryClass = {astro-ph.CO},
       adsurl = {https://ui.adsabs.harvard.edu/abs/2023AJ....165..253H}
}

@ARTICLE{Zhou2023,
       author = {{Zhou}, Rongpu and others},
        title = "{Target Selection and Validation of DESI Luminous Red Galaxies}",
      journal = {The Astronomial Journal},
         year = 2023,
        month = feb,
       volume = {165},
       number = {2},
          eid = {58},
        pages = {58},
          doi = {10.3847/1538-3881/aca5fb},
archivePrefix = {arXiv},
       eprint = {2208.08515},
 primaryClass = {astro-ph.CO},
       adsurl = {https://ui.adsabs.harvard.edu/abs/2023AJ....165...58Z}
}

@ARTICLE{Raichoor2023,
       author = {{Raichoor}, Anand and others},
        title = "{Target Selection and Validation of DESI Emission Line Galaxies}",
      journal = {The Astronomial Journal},
         year = 2023,
        month = mar,
       volume = {165},
       number = {3},
          eid = {126},
        pages = {126},
          doi = {10.3847/1538-3881/acb213},
archivePrefix = {arXiv},
       eprint = {2208.08513},
 primaryClass = {astro-ph.CO},
       adsurl = {https://ui.adsabs.harvard.edu/abs/2023AJ....165..126R}
}

@ARTICLE{Chaussidon2023,
       author = {{Chaussidon}, E. and others},
        title = "{Target Selection and Validation of DESI Quasars}",
      journal = {The Astronomial Journal},
         year = 2023,
        month = feb,
       volume = {944},
       number = {1},
          eid = {107},
        pages = {107},
          doi = {10.3847/1538-4357/acb3c2},
archivePrefix = {arXiv},
       eprint = {2208.08511},
 primaryClass = {astro-ph.CO},
       adsurl = {https://ui.adsabs.harvard.edu/abs/2023ApJ...944..107C}
}



\end{document}